\documentclass[useAMS,usenatbib]{mnras}
\usepackage[normalem]{ulem}

\usepackage{tabularx,ragged2e,booktabs}
\usepackage{natbib}
\usepackage{graphicx}
\usepackage{amsmath,amsfonts,amssymb}
\usepackage{multirow}
\usepackage{hyperref}
\usepackage{color}
\usepackage{cases}
\usepackage[figure]{hypcap}
\usepackage{bm}
\usepackage{capt-of}

\hypersetup{colorlinks, citecolor=blue, pdfduplex=DuplexFlipLongEdge}
\hypersetup{pdftitle=MADSimulationsM87, pdfauthor=Andrew Chael, pdfsubject=Astrophysics}
\def\sgraa{{Sgr~A$^{\ast} $}}
\def\adi{\Gamma_\mathrm{gas}}

\newcommand{\msun}{M_{\odot}}

\def\lsim{\mathrel{\raise.3ex\hbox{$<$\kern-.75em\lower1ex\hbox{$\sim$}}}}
\def\gsim{\mathrel{\raise.3ex\hbox{$>$\kern-.75em\lower1ex\hbox{$\sim$}}}}
\def\gtwid{\mathrel{\raise.3ex\hbox{$>$\kern-.75em\lower1ex\hbox{$\sim$}}}}
\def\proptwid{\mathrel{\raise.3ex\hbox{$\propto$\kern-.75em\lower1ex\hbox{$\sim$}}}}

\title[Two-temperature MAD simulations of M87]{Two-temperature, Magnetically Arrested Disc simulations of the jet from the supermassive black hole in M87}
\author[A. Chael et al.]
{Andrew Chael$^{1}$\thanks{\hbox{E-mail: achael@cfa.harvard.edu}},
Ramesh Narayan$^{1}$
and Michael D.\ Johnson$^{1}$
\\
$^{1}$Harvard-Smithsonian Center for Astrophysics, 60 Garden Street, Cambridge, MA 
02138, USA\\
}

\begin{document}
\maketitle

\begin{abstract}
We present two-temperature, radiative general relativistic magnetohydrodynamic simulations of Magnetically Arrested Discs (MAD) that launch powerful relativistic jets. The mass accretion rates of our simulations are scaled to match the luminosity of the accretion flow around the supermassive black hole in M87.  We consider two sub-grid prescriptions for electron heating: one based on a Landau-damped turbulent cascade, and the other based on heating from trans-relativistic magnetic reconnection. The simulations produce jets with power on the order of the observed value for M87. Both simulations produce spectra that are consistent with observations of M87 in the radio, millimetre, and submillimetre. Furthermore, the predicted image core-shifts in both models at frequencies between 15 GHz and 86 GHz are consistent with observations. At 43 and 86 GHz, both simulations produce wide opening angle jets consistent with VLBI images. Both models produce 230~GHz images with distinct black hole shadows that are resolvable by the Event Horizon Telescope (EHT), although at a viewing angle of $17^\circ$, the 230 GHz images are too large to match EHT observations from 2009 and 2012. The 230 GHz images from the simulations are dynamic on time-scales of months to years, suggesting that repeated EHT observations may be able to detect the motion of rotating magnetic fields at the event horizon. 
\end{abstract}


\section{Introduction}
\label{sec::intro}

The core of the giant elliptical galaxy Messier 87 (M87) launches a relativistic jet that extends for several kiloparsecs \citep{Curtis18}. The jet is dynamic, and it has been observed in wavelengths from radio to $\gamma$-ray \citep{Abramowski2012}. Very-long-baseline interferometry (VLBI) observations at radio and millimetre wavelengths \citep[e.g.][]{Palmer67,Reid1982,Junor1999,Kovalev2007,Ly2007,Asada2012,Hada2016,Mertens2016,Walker2018,KimM87} show that the jet remains collimated on sub-parsec scales into the heart of the galaxy, where it terminates in a bright radio core. 
The radio core has a frequency-dependent position, indicative of self-absorbed emission from the jet that becomes increasingly transparent at higher frequencies \citep{BK79}. At frequencies higher than ${\sim}86$ GHz, the radio core is coincident with a supermassive black hole \citep[SMBH:][]{Hada2011}. The black hole mass has been measured to be $6.6\times10^9 M_\odot$ from stellar dynamics in the surrounding galaxy nucleus (assuming a distance of $D=17.9$ Mpc; \citealt{Gebhardt11}). The mass measured from gas dynamics is a factor of two smaller \citep{Walsh13}. 

Making images of the horizon-scale structure around the black holes in M87 and the Galactic Centre (\sgraa) is a primary objective of the Event Horizon Telescope (EHT), a global millimetre VLBI array with an imaging resolution of ${\approx}15\,\mu$as at 230~GHz \citep{Doeleman09}. The EHT has constrained the compact structure in the core of M87 to the scale of a few Schwarzschild radii \citep{Doeleman12, Akiyama15}.  Assuming a distance to the black hole of $D = 16.7$ Mpc \citep{Mei2007} and a black hole mass of $6.2\times10^9 M_\odot$\citep[][scaled for this distance]{Gebhardt11}, M87's black hole ``shadow''\citep[i.e., the lensed photon orbit;][]{Bardeen,Falcke2000} should have a diameter of ${\sim}40\,\mu$as, making it accessible to imaging by the full EHT \citep{Lu2014, akiyama_M87Imaging, Chael18_Closure}.

The core of M87 is a Low Luminosity Active Galactic Nucleus (LLAGN),  with a luminosity many orders of magnitude below the Eddington limit. LLAGN like M87 constitute the bulk of the SMBH population in the local universe \citep{GreeneHo07, Ho08}. Their low luminosity is naturally produced by an Advection-Dominated Accretion Flow around the central black hole (ADAF: \citealt{Ichimaru77, Rees1982, Abram95, NarayanYi_a,NarayanYi_b, Maha98,Blandford99}). In contrast to the optically thick, geometrically thin accretion discs found in bright AGN \citep{ShakuraSunyaev}, ADAFs in LLAGN are optically thin and geometrically thick (see \citealt{Yuan14} for a review).  Jets like that in M87 are likely powered by the black hole's rotational energy, which is extracted by ordered magnetic fields threading the black hole event horizon \citep{BZ,Tchekhovskoy11,Zaman2014}. 

The dynamics of hot accretion flows and the formation of jets have been extensively investigated using grid-based general relativistic magnetohydrodynamic (GRMHD) and radiative magnetohydrodynamic (GRRMHD) codes \citep[e.g.][]{Komissarov99,DeVilliers03b,Gammie03,Tchekhovskoy10,Tchekhovskoy11,KORAL13,KORAL14,McKinney14,Ryan15,White16,Chandra17}. 
These simulations have demonstrated that jets powered by the black hole spin can be launched from thick discs and accelerated to high Lorentz factors \citep{McKinney06,Komissarov07,McKinney09, Liska2018}. For the specific case of M87, \citet{Dexter12}, \citet{Moscibrodzka_16M87,Moscibrodzka_18}, and \citet{RyanM87} have investigated the spectra and 230 GHz images predicted from various GRMHD simulations.

In hot, low-density, low-accretion rate flows like that around M87 (and \sgraa), Coulomb coupling
between electrons and ions is inefficient  \citep{Maha98}. Electrons and ions will not be in thermal equilibrium with each other, and in most cases the temperature of the single fluid evolved by a GRMHD or GRRMHD simulation will be dominated by the ion temperature, leaving the temperature of the emitting electrons unconstrained. In a simulation, the electron-to-gas
temperature ratio is typically set manually during postprocessing (e.g \citealt{Shch_12, Moscibrodzka_14,Chan_15a} for \sgraa\ and \citealt{Dexter12,Moscibrodzka_16M87} for M87). In this approach, the electron temperature is usually adjusted to find the best fit to the measured spectrum. Nonthermal particle distributions can also be added \citep[e.g.][]{Dexter12, Ball_16, Davelaar_17}. 

Another approach is to \emph{self-consistently} evolve ions and electrons, each with its own thermodynamics and interactions. This approach has been pursued with several different codes \citep{Ressler15, KORAL16, Ryan17}, all of which  evolve a thermal electron population alongside the other standard GRMHD or GRRMHD fluid variables. These methods have been used in several recent simulations of \sgraa\ \citep{Ressler17, Chael18} and of M87 \citep{RyanM87}. 

A substantial source of uncertainty in all these simulations 
arises from the fact that the physical processes that govern viscous dissipation occur on scales much smaller than the finest scales resolved by the grid. The full range of physical processes that govern dissipation at the smallest scales in hot accretion flows is still unconstrained. In \citet{Chael18}, we investigated two candidates for the origin of viscous dissipation in simulations of \sgraa: the Landau-damped turbulent cascade heating prescription of \citet{Howes10} and a prescription for magnetic reconnection heating obtained from particle-in-cell (PIC) simulation data  \citep{Rowan17}. We showed that predictions for the image morphology of \sgraa at 230 GHz and longer wavelengths depend strongly on the choice of heating prescription. The turbulent cascade prescription results in a natural ``disc-jet'' image morphology at longer wavelengths, while the reconnection prescription produces isotropic, disc-dominated emission \citep{Chael18, Issaoun18}.

Recently, \citet{RyanM87} carried out axisymmetric simulations of M87 with two-temperature evolution and frequency-dependent radiative transport. They found that, because M87 is more radiatively efficient than \sgraa, including radiation in the simulation along with temperature evolution of the electrons is critical. They performed simulations at both low ($3.3\times10^{9}\,\msun$) and high ($6.2\times10^{9}\,\msun$)values of the SMBH mass and found that the accretion flow in the high mass, high spin case produced a 230 GHz image consistent with published EHT observations \citep{Doeleman12,Akiyama15}. However, their simulations were performed with weak values of magnetic flux threading the horizon. As a result, the jets in their simulations had a narrower opening angle than that observed in VLBI images of M87, and the jet power was lower than the measured value by several orders of magnitude.

Magnetically Arrested Discs \citep[MADs;][]{Bisnovatyi1976,NarayanMAD} represent the opposite limit to the weak magnetic flux mode of black hole accretion explored in \citet{RyanM87}. In these systems, coherent magnetic flux builds up on the black hole, both launching powerful jets and limiting accretion via magnetic pressure. In GRMHD simulations \citep{Igumenschchev2003,Tchekhovskoy11,McKinney2012,Narayan2012,Sadowski2013}, MADs launch jets powered by the black hole spin with wide opening angles and large jet powers. The measured jet power of M87 is large \citep[${\sim}10^{43}-10^{44}$ erg s$^{-1}$\,;][]{Reynolds96_2,Owen00,Stawarz06,deGasperin2012}, and the jet is launched with a wide opening angle \citep[${\sim}55^\circ$ at 43 GHz ;][]{Walker2018}. Thus, there is reason to suspect that M87 has a magnetically arrested disc at its core.

In this paper we present the results of two fully 3D, two-temperature GRRMHD simulations of Magnetically Arrested Discs around the black hole in  M87 performed using the code \texttt{KORAL} \citep{KORAL13,KORAL14,KORAL16}. In these simulations, we again compare the Landau-damped turbulent cascade heating prescription from \citet{Howes10} with the magnetic reconnection prescription from  \citet{Rowan17, Chael18}. Both simulations are performed assuming a SMBH mass of $6.2\times10^9 M_\odot$ (\citealt{Gebhardt11}, scaled for a distance of 16.7 Mpc; \citealt{Mei2007}) and a dimensionless spin of $a=0.9375$. These simulations are the first Magnetically Arrested Discs evolved with two-temperature electron-ion thermodynamics.

In Section~\ref{sec::phys}, we briefly review the method used in \citet{KORAL16} and \citet{Chael18} for evolving
thermal electron and ion entropies, as well as the two heating prescriptions considered in this paper. In Section~\ref{sec::num}, we describe the setup of our simulations and our radiative transfer method for producing images and spectra. We present our results in Section~\ref{sec::results}. In Section~\ref{sec::properties}, we first discuss the time-averaged characteristics of our models, including the accretion rate, jet power, and distributions of temperature, magnetic field, and radiation. We then present spectra for our simulations in Section~\ref{sec::spectra} and discuss variability in the 230 GHz light curves in Section~\ref{sec::variability}. In Section~\ref{sec::images}, we show images at various millimetre wavelengths and compare to VLBI images (Section~\ref{sec::43}) and core-shift measurements (Section~\ref{sec:core_shift}). We present 230 GHz images and compare to previous EHT measurements in Section~\ref{sec::230GHz}. We discuss the 230 GHz image structure and time evolution, and we explore which regions of the accretion flow are responsible for the image features. In Section~\ref{sec::discussion} we compare our results with other models and discuss their implications for Event Horizon Telescope observations and future work. We conclude in Section~\ref{sec::summary}.

\section{Equations}
\label{sec::phys}
\subsection{Evolution Equations}
\begin{figure*}
\centering
\includegraphics*[width=0.99\textwidth]{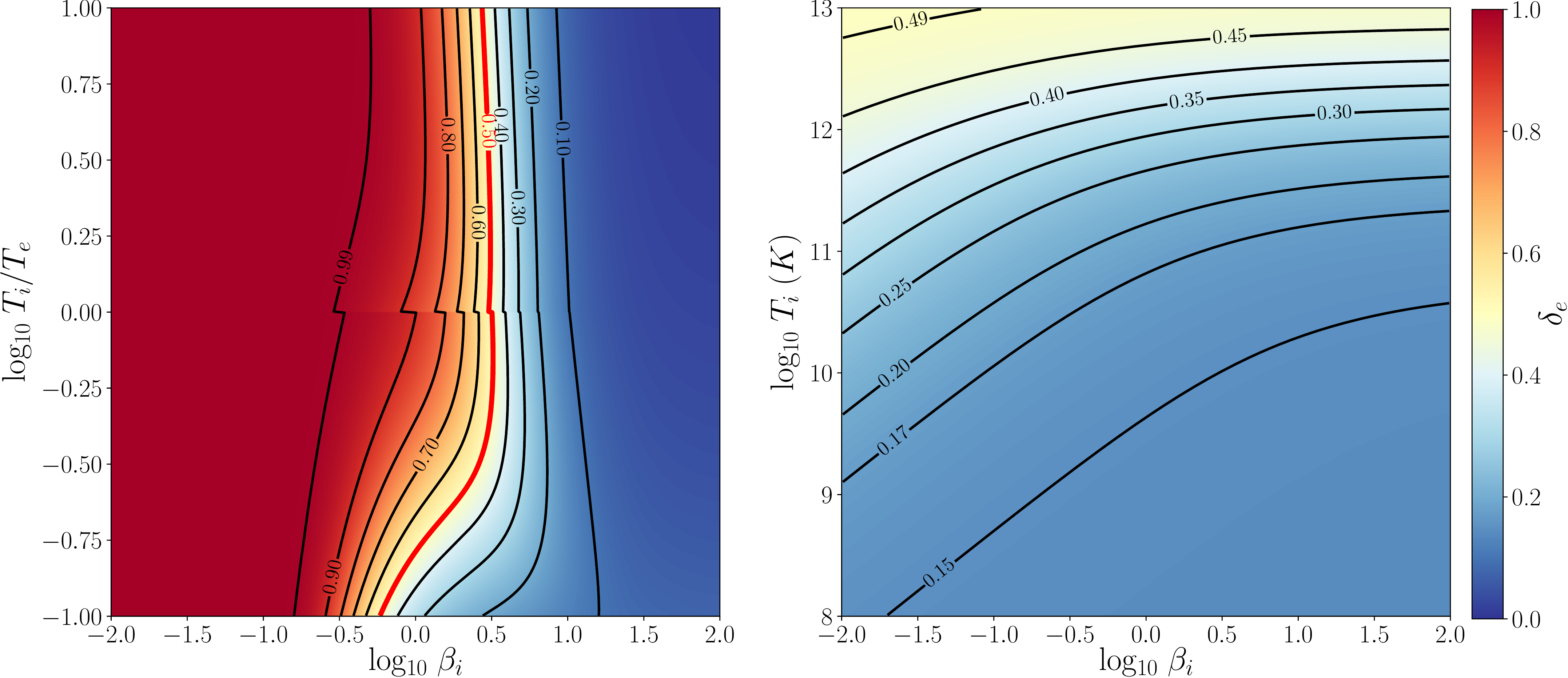}
\caption{The two prescriptions for electron heating used in this work, taken from figs. 1 and 3 of \citet{Chael18}. (Left) Turbulent cascade heating prescription \citep{Howes10}. The electron heating fraction $\delta_\mathrm{e}$ is shown as a function of the plasma-beta parameter $\beta_\mathrm{i}$ and the temperature ratio $T_\mathrm{i}/T_\mathrm{e}$. This prescription transitions rapidly from putting most of the dissipated energy into electrons at low $\beta_\mathrm{i}$ to putting most of the dissipated energy into ions at high $\beta_\mathrm{i}$. The red contour denotes $\delta_\mathrm{e} = 0.5$. (Right) Reconnection heating prescription \citep{Rowan17,Chael18} obtained by fitting to PIC simulation data. For clarity, we assume that $T_\mathrm{i}=T_\mathrm{e}$ and plot $\delta_\mathrm{e}$ as a function of $T_\mathrm{i}$ and $\beta_\mathrm{i}$ instead of the PIC simulation variables $\beta_\mathrm{i}$  and $\sigma_w$. In this prescription, $\delta_\mathrm{e}$ never exceeds 0.5. 
}
\label{fig::heating}
\end{figure*}

In this section, we briefly review the method used in the GRRMHD code \texttt{KORAL} to evolve a two-temperature fluid (\citealt{KORAL16}, also summarized in \citealt{Chael18}). 

We consider two fluids in spacetime, electrons and ions (assumed for the remainder of this work to be entirely ionized Hydrogen). Charge neutrality demands these fluids have the same number density $n$ and four velocity $u^{\mu}$ everywhere, but without efficient processes to bring them into equilibrium they can have distinct thermal energy densities $u_{\rm e} \neq u_{\rm i}$ and temperatures $T_{\rm e} \neq T_{\rm i}$. 

Together, electrons and ions form a single mixed fluid, which is characterized by a mass density dominated by the ions,
$\rho = m_p n$, and a total internal energy density $u = u_{\rm e} + u_{\rm i}$.  
The total pressure $p = p_{\rm e} + p_{\rm i}$ can then be expressed with an effective adiabatic index; $\adi$, $p=(\adi-1)u$. In a hot accretion flow, which typically has temperatures $>10^{10}$ K in the inner regions, electrons become relativistic and their adiabatic index can decrease from $5/3$ towards $4/3$. In the innermost regions where $T_{\rm i}\rightarrow10^{11}$ K, even the ions become quasi-relativistic. Thus, the effective gas adiabatic index $\adi$ takes on values in the range $4/3 \le \adi \le 5/3$ depending on the local temperatures and energy densities of the two component species (see \citealt{KORAL16} for the form of the adiabatic indices as a function of temperature used in \texttt{KORAL}).

The MHD stress-energy tensor $T^\mu_{\;\;\nu}$ consists of contributions from the fluid variables as well as the magnetic field four-vector $b^\mu$ \citep{Gammie03}:\footnote{The magnetic field strength in Gauss is $|B| = \sqrt{4\pi b^{\mu}b_{\mu}}$}
\begin{equation}
 \label{eq::tmunu}
 T^\mu_{\;\;\nu} = \left(\rho + u + p + b^2\right)u^\mu u_\nu + \left(p + 
\frac{1}{2}b^2\right)\delta^\mu_{\;\;\;\nu} - b^\mu b_\nu.
\end{equation}

\texttt{KORAL} treats the frequency-integrated radiation field $R^\mu_{\;\;\nu}$ as a second perfect fluid. The radiation fluid is described
by its rest frame energy density $\bar{E}$ and its four-velocity $u^\mu_R \neq u^\mu$:
\begin{equation}
 R^\mu_{\;\;\nu} = \frac{4}{3}\bar{E}u^\mu_R u^{\vphantom{\mu}}_{R\,\nu} + 
\frac{1}{3}\bar{E}\delta^\mu_{\;\;\nu}.
\end{equation}
In this formulation (M1 closure), radiation is described at each spacetime point by four bolometric quantities: $\bar{E}$ and the three-velocity $u^i_R$. We also track a fifth quantity, the photon number density $\bar{n}_R$, which encodes information about the mean photon frequency $\bar{E}/h\bar{n}_R$. In contrast to this frequency-integrated approach, \citet{Ryan15,Ryan17,RyanM87} use a Monte-Carlo approach which represents the radiation field with many individual particle ``superphotons'' with different frequencies that are emitted and absorbed in between the fluid evolution timesteps. 

The set of GRRMHD equations for evolving the total fluid, the magnetic field, the frequency-integrated radiation field, and the photon number are \citep{Gammie03, KORAL14, Compt15}
\begin{align}
\label{eq::GRRMHD}
(\rho u^\mu)_{;\mu} & = 0, \\
T^{\mu}_{\;\;\nu;\mu} &= G_\nu, \\
R^{\mu}_{\;\;\nu;\mu} &= -G_\nu, \\
\label{eq::photev}
(\bar{n}^{\vphantom{\mu}}_R u^\mu_{R})_{;\mu} &= \dot{\bar{n}}_R , \\
\label{eq::GRRMHD4}
F^{*\mu}_{\;\;\;\;\nu;\mu} &=0, 
\end{align}
where $F^{*\mu\nu} = b^\mu u^\nu - b^\nu u^\mu$ is the dual of the MHD Maxwell tensor, $G_\nu$ is the four-force density that couples the radiation and gas (see \citealt{KORAL16} for the precise form), and $\dot{\bar{n}}_R$ is the frame-invariant photon production rate 
(see \citealt{Compt15}).

In evolving electrons and ions, we consider the entropy per particle of each, $s_\mathrm{e}$ and $s_\mathrm{i}$.  The temperatures $T_\mathrm{e}$, $T_\mathrm{i}$ and energy densities $u_\mathrm{e}$, $u_\mathrm{i}$ are functions of the species entropy and number density (see \citealt{KORAL16} and the Appendix of \citealt{Chael18}). The species entropies are evolved using the first law of thermodynamics: 

\begin{align}
\label{eq::ent_ev}
T_\mathrm{e}\left(n s_\mathrm{e} u^\mu\right)_{;\mu} &= \delta_\mathrm{e} q^{\rm v} + 
q^{\rm C} - \hat{G}^0,  \\
\label{eq::ent_ev2}
T_\mathrm{i}\left(n s_\mathrm{i} u^\mu\right)_{;\mu} &= (1-\delta_\mathrm{e}) q^{\rm v} - q^{\rm C},
\end{align}
where $q^\mathrm{v}$ is the dissipative heating rate, $\delta_\mathrm{e}$ is the fraction of the dissipative heating that goes into electrons, $q^\mathrm{C}$ is the energy exchange rate from ions to electrons due to Coulomb coupling \citep{Stepney83}, and $\hat{G}^0$ is the radiative cooling rate. 

The physical processes that produce dissipation  occur at scales far smaller  than the simulation grid. We identify the \emph{total} dissipative  heating $q^{\rm v}$ numerically by evolving the thermal entropies adiabatically over a timestep $\Delta \tau$. We then compare the sum of the adiabatically
evolved energy densities, $u_{\rm i,\,adiab}$ and $u_{\rm e,\,adiab}$, to the separately-evolved total gas energy $u$, thereby estimating the dissipative heating in the total fluid:
\begin{equation}
 \label{eq::vischeat}
  q^{\rm v} = \frac{1}{\Delta \tau}\left[u - u_{\rm i,\,adiab} - u_{\rm e,\,adiab}\right].
\end{equation}
The \emph{fraction} $\delta_\mathrm{e}$ of the heating that goes into the electrons, however,
must be determined by a sub-grid prescription.

\subsection{Electron Heating Prescriptions}
\label{sec::heating}

In this work, we again consider the two sub-grid electron heating prescriptions for $\delta_{\rm e}$ that we previously applied to simulations of \sgraa\ \citep[for a fuller discussion of the physics behind these prescriptions, see][]{Chael18}. These prescriptions depend on three parameters: the ``plasma-beta'' $\beta_\mathrm{i}$, the magnetization $\sigma_\mathrm{i}$, and the temperature ratio $T_\mathrm{e}/T_\mathrm{i}$.

The plasma-beta parameter $\beta_\mathrm{i}$  is the ratio of the thermal ion pressure to the magnetic pressure:
\begin{equation}
 \label{eq::beta}
 \beta_\mathrm{i} = \frac{8\pi \, n_\mathrm{i} k_{\rm B} T_\mathrm{i}}{|B|^2},
\end{equation}
and the magnetization $\sigma_\mathrm{i}$ compares the magnetic energy density to the rest-mass energy density of the fluid:
\begin{equation}
 \label{eq::sigmai}
 \sigma_\mathrm{i} = \frac{|B|^2}{4\pi \, n_\mathrm{i} m_\mathrm{i} c^2}.
\end{equation}
While in general $\sigma_\mathrm{i} \ll 1$ in the disc, in our MAD simulations $\sigma_\mathrm{i}$ exceeds unity in the jet as well as close to the black hole. 

Both heating prescriptions $\delta_{\rm e}$ we consider are plotted as a function of different plasma parameters in Fig.~\ref{fig::heating}. The first prescription for $\delta_{\rm e}$ is taken from calculations of the Landau damping of a MHD turbulent cascade in a weakly collisional plasma \citep{Howes10}. This prescription is based on nonrelativistic calculations with $\sigma_\mathrm{i} \ll 1$  \citep{Howes2008b, Howes2008a}, and while it matches solar wind measurements \citep{Howes11}, it may not be well-adapted to relativistic systems like the M87 accretion flow.  Recently, however, \citet{Kawazura18} performed numerical simulations of the turbulent damping process that indicate the qualitative behavior of the \citet{Howes10} prescription holds even in relativistic turbulent plasmas. 

The \citet{Howes10} turbulent cascade prescription is a function of the temperature ratio and $\beta_\mathrm{i}$. It most strongly depends on $\beta_\mathrm{i}$, sharply transitioning from primarily heating electrons ($\delta_\mathrm{e} \approx 1$) at low $\beta_\mathrm{i}$, to primarily heating ions ($\delta_\mathrm{e} \approx 0$) at high $\beta_\mathrm{i}$. While in general we expect radiation to cool electrons to lower temperatures than ions, in \citet{Chael18} we saw that using this prescription nonetheless results in an electron temperature that can greatly exceed the ion temperature in the jet region (where $\beta_\mathrm{i}\ll1$). 

The \citet{Howes10} electron heating prescription is based on a model of plasma turbulence truncated by Landau damping of turbulent eddies at small scales.  Another model for turbulent dissipation suggests that MHD turbulence may instead be truncated at small scales by magnetic reconnection, as eddies become sheet-like and fragment via the tearing instability \citep{Carbone1990, Boldyrev2006, Boldyrev2017, Loureiro2017}. This  model of the dissipation scale of plasma turbulence motivates our second heating prescription, obtained by measuring electron heating in particle-in-cell (PIC) simulations of trans-relativistic reconnection.
\citet{Rowan17} explored magnetic reconnection with PIC simulations in plasmas over a trans-relativistic range of  temperatures and magnetic field strengths (see also \citealt{Werner2018} for a similar study). A fitting function to $\delta_e$ as measured in the reconnection PIC simulations of \citet{Rowan17} was presented in \citet{Chael18}. Note that the simulations in \citet{Rowan17} had no guide field perpendicular to the reconnection; the efficiency of electron heating in the strong guide field regime is yet to be studied in detail and may be qualitatively different than in the zero guide field case.

The reconnection prescription differs substantially from the turbulent cascade prescription. At a fixed temperature, decreasing $\beta_{\rm i}$ results in energy being shared equally between the species downstream from the reconnection region. Thus, $\delta_e$ approaches a maximum of 0.5  in the most magnetically dominated plasmas. We should thus never see $T_\mathrm{e}>T_\mathrm{i}$ when using the reconnection prescription. At the other limit of large $\beta_{\rm i}$, $\delta_{\rm e}$ approaches a constant value that depends only on $\sigma_w$ (defined with respect to the fluid enthalpy $w$);\footnote{
$\sigma_{w}$ and $\beta_{\rm i}$ are the independent variables in the PIC simulations of \citet{Rowan17}. See Section 2.2 of \citet{Chael18} for a discussion of the difficulties of simply interpreting $\sigma_w$ in terms of $\sigma_{\rm i}$ in relativistic plasmas.} $\delta_{\rm e}$ is nonzero even for $\sigma_w < 10^{-3}$. This behavior is qualitatively different from the prescription of \citet{Howes10}, which sends $\delta_{\rm e}\rightarrow 0$ for similarly large values of $\beta_{\rm i}$. In the \sgraa\ simulations of \citet{Chael18}, the floor on $\delta_{\rm e}$ in the reconnection prescription resulted in hotter discs than in the simulations that used the turbulent prescription. 

\section{Numerical Simulations}
\label{sec::num}

\subsection{Units}
\label{sec::units}

In both simulations presented in this work, we fix the distance to M87 as $D = 16.7$ Mpc \citep{Mei2007} and fix the  
black hole mass to $6.2\times10^9 M_\odot$ \citep[][scaled for this distance]{Gebhardt11}. We take the dimensionless black hole spin in both simulations as $a=0.9375$.

For this mass, the gravitational length scale of M87 is  
$r_{\rm g} = GM/c^2 = 9.2\times10^{14} \, \text{cm} = 61 \,\text{AU}$. The corresponding angular scale is $r_{\rm g}/D = 3.7\,\mu$as. The gravitational time-scale is $t_{\rm g}=r_{\rm g}/c=3\times10^4\,\text{s}=8.5\,\text{hr}$. 

M87's Eddington luminosity is $L_\mathrm{Edd}=7.8\times10^{47}$ erg~s$^{-1}$. The Eddington accretion rate is $\dot{M}_\mathrm{Edd}=L_\mathrm{Edd}/\eta c^2=77 \, M_\odot$ yr$^{-1}$, where for our chosen value of spin, we set the efficiency $\eta=0.18$, as expected for a thin accretion disc with $a=0.9375$ \citep{NovikovThorne}. 

\subsection{Simulation Setup}
\label{sec::setup}

Our simulations were performed in the Kerr metric using a modified Kerr-Schild coordinate grid that is exponential in radius and concentrates grid cells near the equator (see the Appendix of \citealt{Chael18} for the transformation between our coordinates and standard Kerr-Schild coordinates). We used a resolution of $288\times224\times128$ cells in the $r$, $\theta$, and $\phi$ directions, respectively, which well-resolves the magnetorotational instability (MRI) and enabling accretion. To capture the evolution of the jet at large radii, we set the outer boundary of the simulation box at $10^5 \, r_{\rm g}$. 

We set up initial equilibrium gas torii using the model of \citet{Penna13}. To build up magnetic field to the point where the disc reaches the saturation value of magnetic flux and becomes magnetically arrested, we initialized the torus with a single weak ($\beta_\text{max} = 100$) magnetic field loop centered around $r\approx50\,r_{\rm g}$. The initial energy in electrons was set at one per cent of the total gas energy, with the remainder in ions. 

\texttt{KORAL} solves Eqs.~\eqref{eq::ent_ev} and \eqref{eq::ent_ev2} for the electron and ion thermodynamics in parallel with the  conservation Eqs.~\eqref{eq::GRRMHD}~--~\eqref{eq::photev} for the matter and radiation fluids and the magnetic field induction equation Eq.~\eqref{eq::GRRMHD4}. The advection of quantities across cell walls is computed explicitly by reconstructing the appropriate fluxes at the cell walls using the second-order piecewise parabolic method (PPM). The source and coupling terms in the evolution equations are then applied implicitly at each cell center using a Newton-Raphson solver \citep{KORAL13,KORAL14,KORAL16}.

Outflowing boundary conditions are used at the inner and outer radial boundaries, and reflecting boundary conditions are imposed at the the polar axes. In the nearest two cells to the polar axis, we control numerical instability from fluid flow across the poles by replacing the value of $u^\theta$ with the value from the third cell at the end of each timestep.

In the jet region, high fluid velocities rapidly evacuate the funnel and cause the fluid density to drop without bound. In order to ensure  the numerical stability of our simulations, we put a global ceiling on the magnetization $\sigma_{\rm i}$, as measured in the zero angular momentum observer (ZAMO) frame \citep{McKinney2012}. In this frame, the fluid density is increased to bring the magnetization back to our chosen limit, $\sigma_\mathrm{i,max} = 100$.

We evolved one simulation using the \citet{Howes10} prescription for dividing viscous dissipation between electrons and ions; we refer to this simulation as \texttt{H10}. We evolved the other simulation using the magnetic reconnection prescription of \citet{Rowan17, Chael18}; we refer to this simulation as \texttt{R17}. We first ran the two simulations for $10^4 \, t_{\rm g}$ in 3D. During this this time, both simulations formed a thick disc at small radii and accumulated magnetic flux on the black hole horizon that exceeds the MAD threshold of ${\sim}50 \, \sqrt{\dot{M}c } \, r_{\rm g}$ \citep{Tchekhovskoy11,McKinney2012}. At this point, we rescaled the gas density by a factor of 1/4 and magnetic field by 1/2 (keeping the temperatures and magnetization fixed), so that the 230 GHz flux density from the models was approximately equal to the $0.98\pm0.04$ Jy  of compact emission measured by the EHT in 2009 and 2012 \citep{Doeleman12,Akiyama15}.

We then ran the simulations from the rescaling point for another $1000\,t_{\rm g}$ to allow the simulations to settle into a new equilibrium. The results in the following Sections were taken from the final $5000\,t_{\rm g}$, from $t=11,000\,t_{\rm g}$ to $t=16,000\,t_{\rm g}$.  The absence of large secular evolution in the accretion rate and 230 GHz light curve as a function of time in both models (Fig.~\ref{fig::acc_mad}) indicates the system has likely settled into a new equilibrium after the rescaling by the time we start our measurements. 

In both simulations, we define the region of inflow equilibrium inside which we expect the fluid quantities to be reasonably converged by finding where the characteristic accretion time $t_{\rm acc} = 5000\,t_{\rm g}$. At a given radius $r$, we define  \citep{Narayan2012}:
\begin{equation}
\label{eq::inflow}
 t_{\rm acc} = \frac{r}{\left|v_r\right|},
\end{equation}
where $v_r = u^r / u^t$ is the Boyer-Lindquist radial three-velocity. 

Over the final $5000\,t_{\rm g}$ period, the inflow equilibrium region in the disc extends to ${\approx}30 \, r_{\rm g}$ (Fig~\ref{fig::sym1}). In the fast moving jet, the region of outflow equilibrium extends to ${\approx}4700 \, r_{\rm g}$ (Fig.~\ref{fig::sym_lorentz}).

In the Appendix, we present results from a convergence test where  we performed the \texttt{R17} simulation at a lower resolution of $192\times128\times64$ cells in radius, polar angle, and azimuth up to $15,000\,t_{\rm g}$. The general agreement between the synchrotron spectrum and average distributions of electron temperature, density, and magnetization at these two resolutions suggests that the fundamental quantities of interest in these simulations are stable with resolution, except in regions very closer to the black  hole where $\sigma\rightarrow\sigma_{\rm  max}=100$.

\subsection{Reliability of Emission from High Magnetization Regions}
\label{sec::reliability}
In highly magnetized regions ($\sigma_\mathrm{i}>1$), the plasma dynamics and thermodynamics in GRMHD simulations become increasingly suspect. Because the magnetic field dominates the energy budget in these regions, small errors in the total energy evolution can induce large changes in the internal energy and plasma temperature. At high $\sigma\gtrsim100$, these errors typically lead the code to crash, as the implicit solver fails to converge on a solution for the internal energy density from the conserved quantities. 

\begin{figure}
\centering
\includegraphics*[width=0.45\textwidth]{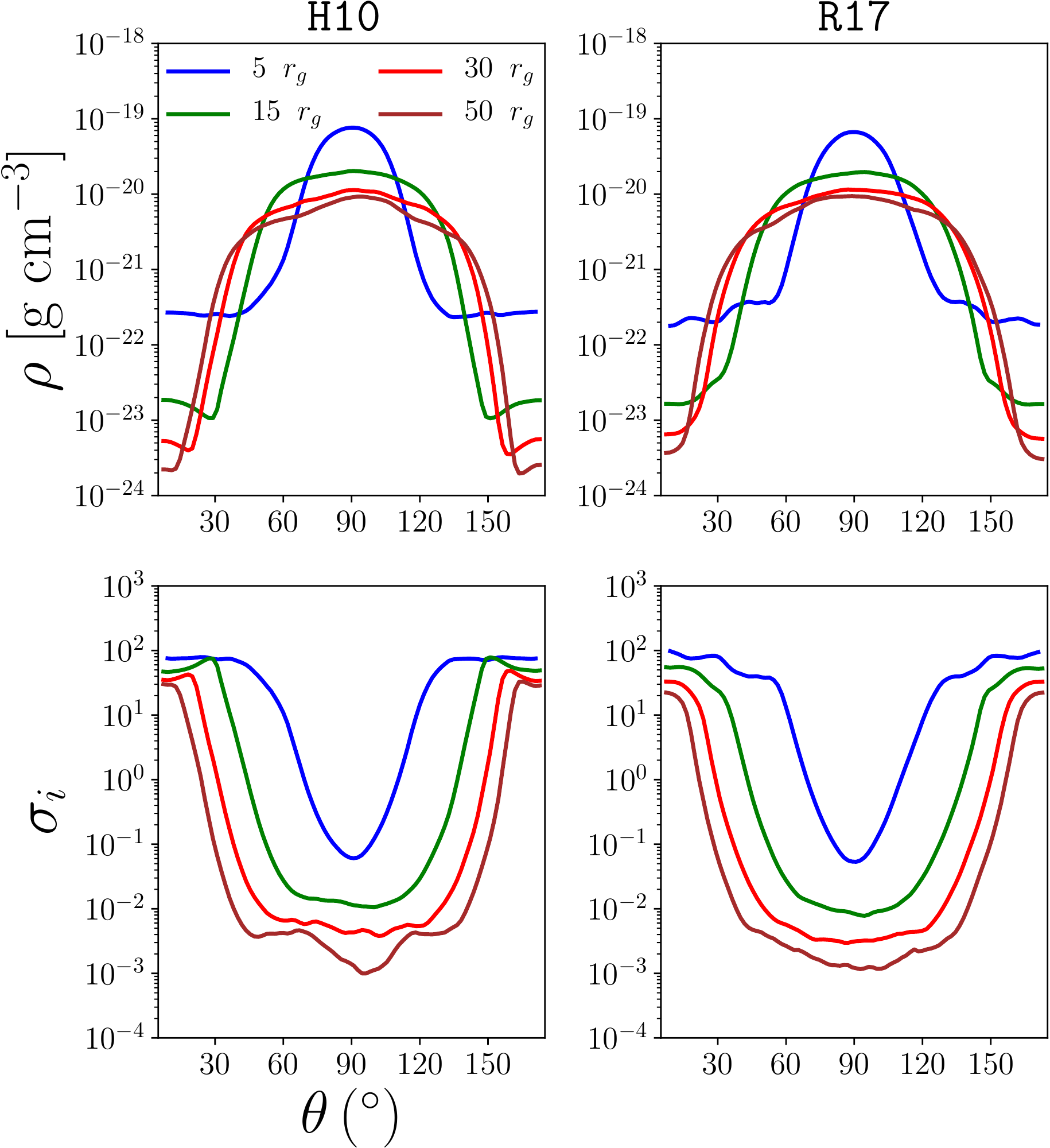}
\caption{Azimuth and time-averaged density (top) and magnetization (bottom) as a function of polar angle $\theta$ for the two simulations at four radii: $r=5 \, r_{\rm g}$ (blue), $r=15 \, r_{\rm g}$ (green),  $r=30 \, r_{\rm g}$ (red), and $r=50 \, r_{\rm g}$ (brown). Snapshot quantities were averaged in azimuth and then time-averaged from $11,000-6,000\,t_{\rm g}$. These data were not symmetrized over the equatorial plane. The ceiling on the magnetization $\sigma_{\rm i,max}=100$ (which we impose in the ZAMO frame) imprints itself as a floor on the density that takes effect at the same polar angle $\theta$. Because the radiation produced in this region is unreliable, we cut regions where $\sigma_{\rm i}>25$ in the radiative transfer computations.
}
\label{fig::symc2}
\end{figure}

As discussed in Section~\ref{sec::setup}, we impose a ceiling on the magnetization $\sigma_{\rm i,max}=100$ to ensure numerical stability. This ceiling results in a constant injection of gas density in the innermost jet regions. Fig.~\ref{fig::symc2} illustrates this ceiling on $\sigma_{\rm i}$ by showing profiles of $\rho$ and $\sigma_{\rm i}$ versus polar angle $\theta$ in the time- and azimuth-averaged data from our simulations. At each radius, the density levels off at a floor value inside the polar angle $\theta$ where $\sigma_{\rm i}$ hits the simulation ceiling $\sigma_{\rm i, max} = 100$. This leveling off is a numerical artifact, and  therefore, radiation from these regions will be artificially intense. We must not include these regions where floors are active in spectra and images generated from the simulations. 

Even in regions where the magnetization is not so strong as to lead to numerical instabilities and the imposition of the density floor, however, it remains a worrying possibility that errors in the thermodynamic evolution may still build up to bias our simulation results. The degree of unreliability of the radiation from the plasma temperature in these regions ($100>\sigma_{\rm i}>1$) is more difficult to assess than in the regions where the density is obviously artificially high (Fig.~\ref{fig::symc2}). This potential unreliability is a problem in all GRMHD simulations (not just two-temperature ones) and in nearly all disc configurations (not just MADs), but it is a particularly significant concern in our case. 

In producing spectra and images from GRMHD simulations, it is standard practice to only consider regions less magnetized than some cutoff value, $\sigma_{\rm i} < \sigma_{\rm cut}$. In most cases, $\sigma_{\rm cut} = 1$ is chosen as a conservative cutoff, eliminating all radiation from all magnetically dominated regions. For non-MAD simulations \citep[e.g.][]{Ressler15,Ressler17,Chael18,RyanM87}, this choice is unlikely to substantially affect the results, as the emission from regions $\sigma_{\rm i} > 1$ is not a significant component the spectrum \citep[see e.g.][Appendix C]{Ressler17}.

In the MAD simulations considered in this work, the primary features of interest -- the jet and near-horizon region -- are highly magnetized. Including at least some emission from the $\sigma_{\rm i}\gtrsim 1$ regions may be necessary to compare our simulations to observations. Consequently, we choose a value of $\sigma_{\rm cut}=25$,  a factor of four lower than the $\sigma_{\rm i}=100$ ceiling where density floors are imposed. This choice eliminates radiation from the density floor regions where the density is not set by mass loading from the disc, and where we know the thermodynamic evolution to be unreliable and unstable (Fig.~\ref{fig::symc2}). We explore the effects of this choice in detail in Section~\ref{sec::sigcut}. We find our results are sensitive to the choice of $\sigma_{\rm cut}$, indicating that it is an important free parameter in considering emission from MAD simulations. Future work on identifying the regions from where emission is reliable in highly-magnetized flows (as well as on more robust methods for evolving thermodynamics in these flows) will be critical in making firm conclusions in comparing images and spectra to data from these sources.  

Finally, we note that the potential unreliability of the thermodynamics outside the bulk of the disc is not even entirely confined to high $\sigma_{\rm i}$ regions.  Along the jet wall, even in regions with $\sigma_{\rm i} < 1 $, the density gradient is large, and there is an effective contact discontinuity between the funnel/``corona'' region and the disc. At this interface, the large entropy and density gradients are difficult for the Riemann solver to handle without substantial diffusion, which leads to non-negligible, time-averaged negative heating rates from Eq.~\eqref{eq::vischeat}\citep{Ressler17}.  This problem may be tractable with extremely high resolution simulations that can resolve this interface (now potentially feasible with GPUs, \citealt{Liska2018}), and using more advanced Riemann solvers than the Lax-Friedrichs solver typically used in GRMHD codes \citep[e.g. the Harten-Lax-van-Leer-Discontinuities solver used in][]{White16}.

\subsection{Radiative Transfer}
\label{sec::radiation}

We produced spectra, images, and lightcurves from our simulations using two post-processing codes. For computing full spectra, we used  
\texttt{HEROIC}, \citep{Zhu_2015, Narayan16_HEROIC}, a code that solves for the spectrum and angular distribution of radiation 
at each grid position self-consistently. Inverse Compton scattering is included along with free-free (from both $\rm{e}-\rm{e}$ and $\rm{e}-\rm{i}$ interactions) and synchrotron emission and absorption. At millimetre and radio wavelengths, synchrotron radiation dominates the emission. To produce high-resolution images and lightcurves in the millimetre and lower frequencies, we used the ray tracing and radiative transfer code \texttt{grtrans} \citep{Dexter16} with thermal synchrotron opacities. We also use \texttt{grtrans} to compute synchrotron-only spectra for $\nu<10^{12}$ Hz.

In both codes, we use the ``fast-light'' approximation, where radiation is produced from each simulation snapshot independently, ignoring the evolution of the fluid as the photons propagate. We used the value $\sigma_{\rm cut} = 25$ throughout (Section~\ref{sec::reliability}) and we did not rescale the density or magnetic field of the models in postprocessing.

The jet inclination angle of M87 is constrained from observed ``super-luminal'' motion of jet components in VLBI images \citep{HeinzBegelman97}. In the present paper, we use an inclination angle of $17^\circ$ \citep{Mertens2016,Walker2018}. We choose to measure this angle up from the lower pole, so that the sense of rotation of the accretion disc and black hole spin is clockwise on the sky. This is the preferred orientation of the jet angular momentum vector as determined by the differential brightening and pattern velocities of the jet limbs in VLBI images \citep{Walker2018}. To match the orientation of the M87 jet on the sky at $-72^\circ$ east of north \citep{Reid1982}, we rotate our computed images $108^\circ$ counterclockwise.

\section{Results}
\label{sec::results}

\begin{figure*}
\centering
\includegraphics*[width=0.99\textwidth]{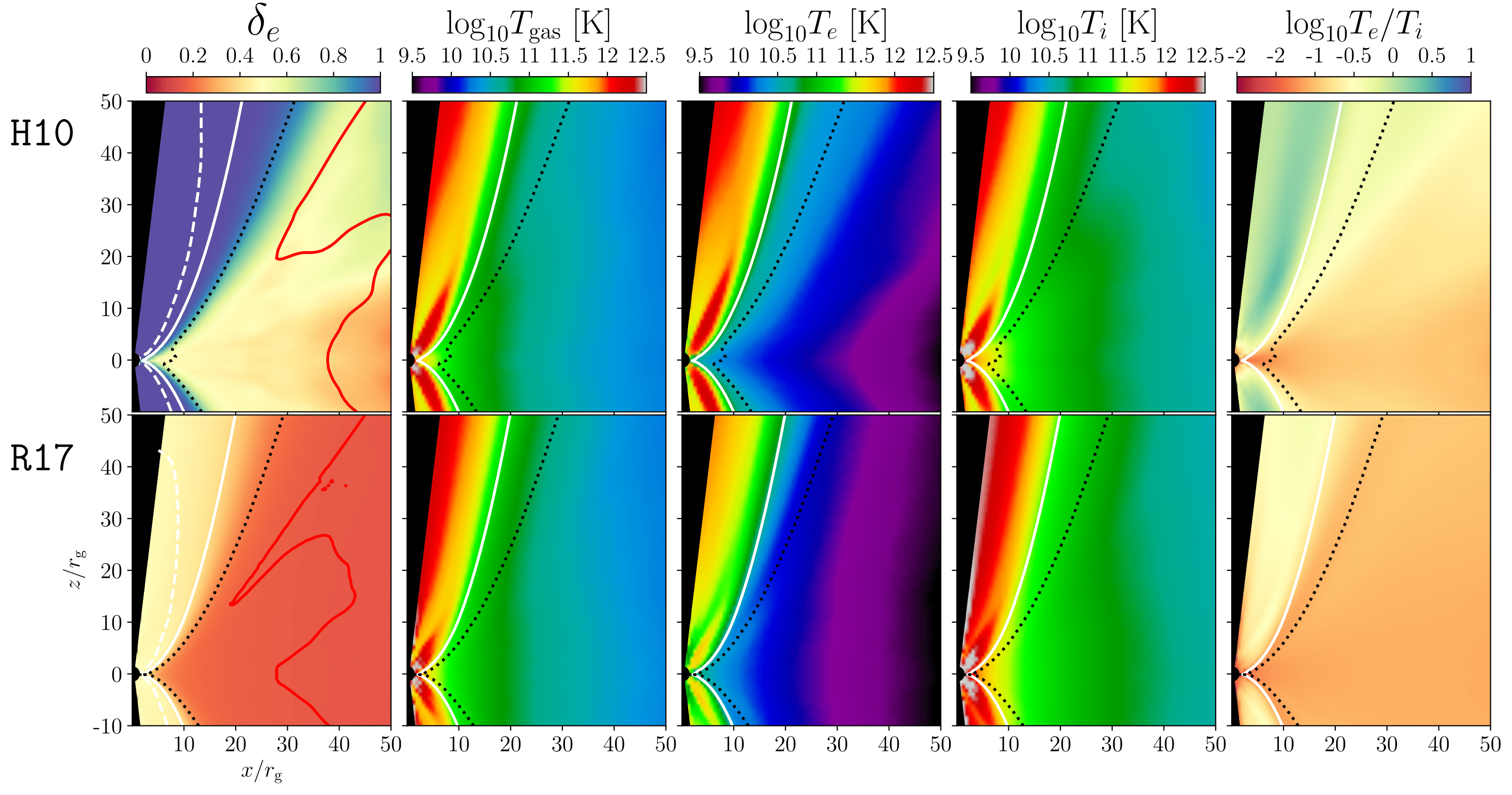}
\caption{
Time- and azimuth-averaged thermodynamic quantities from the two simulations over the period $t=11,000-16,000\,t_{\rm g}$. The top row shows quantities for the model \texttt{H10} heated by the turbulent cascade prescription, and the bottom row shows quantities for the model \texttt{R17} heated by magnetic reconnection. Snapshot quantities were averaged in azimuth and then time-averaged for $5,000 \, t_{\rm g}$. The resulting averages were symmetrized over the equatorial plane.
From left to right, the quantities shown are the electron heating fraction $\delta_\mathrm{e}$, the combined gas temperature $T_{\rm gas}$ in K, 
the electron temperature $T_\mathrm{e}$, the ion temperature $T_\mathrm{i}$, and the electron-to-ion temperature ratio $T_\mathrm{e}/T_\mathrm{i}$. The solid white contour in each panel denotes the surface where $\sigma_{\rm i}$=1, and the dashed black contour shows the surface where the Bernoulli parameter (Eq.~\ref{eq::bernoulli}) ${\rm Be} = 0.05$, which we take as the definition of the jet-disc boundary. The solid red contour in the first column indicates the boundary of the inflow equilibrium region, defined such that $t_{\rm acc} = 5000\,t_{\rm g}$ (Eq.~\ref{eq::inflow}). The dashed white contour in the third panel shows the $\sigma_{\rm i}$=25 surface; this is the maximum $\sigma_{\rm i}$ included in the radiative transfer (see Section~\ref{sec::radiation}) 
}
\label{fig::sym1}

\centering
\includegraphics*[width=0.99\textwidth]{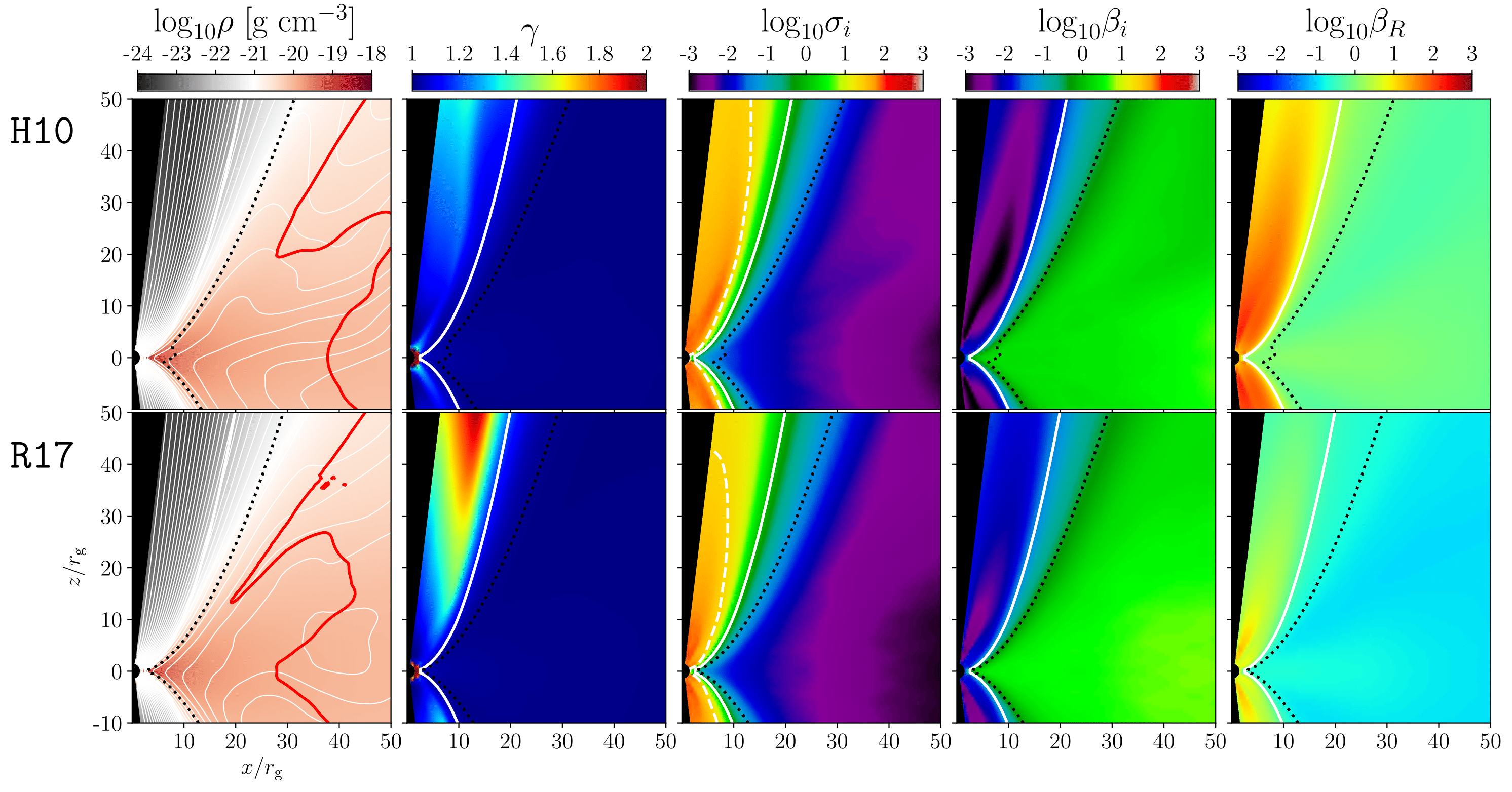}
\caption{
Additional time- and azimuth-averaged properties of the two simulations.  From left to right, the quantities displayed are the density $\rho$ in g cm$^{-3}$, the bulk Lorentz factor $\gamma$, the plasma magnetization $\sigma_\mathrm{i}$,  the ratio of ion thermal pressure to magnetic pressure $\beta_\mathrm{i}$, and the ratio of radiation pressure to thermal pressure $\beta_R$. In the first column, white contours show the poloidal magnetic field in the averaged data. 
}
\label{fig::sym2}
\end{figure*}

\begin{figure*}
\centering
\includegraphics*[width=.95\textwidth]{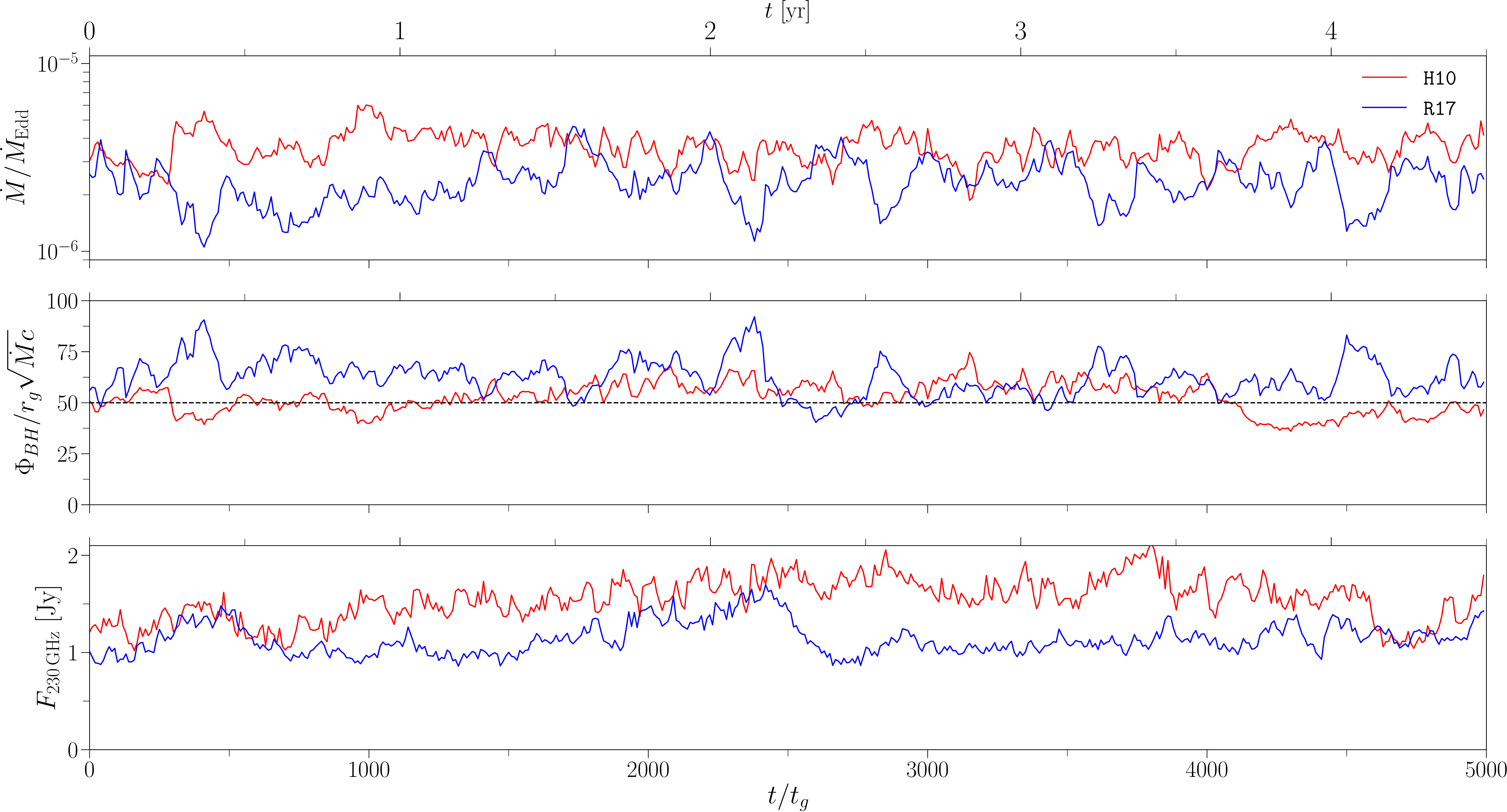}
\caption{Time variability of the two MAD simulations \texttt{H10} (red) and \texttt{R17} (blue) plotted over the 5000 $t_{\rm g} = 4.84 $ yr period from $11,000\,t_{\rm g}$ to  $16,000\,t_{\rm g}$. (Top) Mass accretion rate $\dot{M}/\dot{M}_{\rm Edd}$. Both simulations show strong variability in the accretion rate as fluid parcels slip through the magnetic-pressure-dominated region close to the black hole horizon. (Middle) The dimensionless MAD parameter representing the amount of magnetic flux threading the black hole. Both models are well in the MAD regime $\Phi_{\rm BH}/\sqrt{\dot{M}c}\, r_{\rm g} \gtrsim 50$ \citep{McKinney2012}, but the reconnection heated model \texttt{R17} has a systematically higher magnetic flux on the horizon for most of the time considered and a correspondingly lower accretion rate, which is suppressed by the additional magnetic pressure. (Bottom) 230 GHz light curves computed from high-resolution \texttt{grtrans} images of the two simulations.
}
\label{fig::acc_mad}
\end{figure*}

\begin{table*}
\centering
\begin{tabular}{l||ll|ll|llll}
\hline 
Model  & Spin & Heating & $\langle \dot{M}/\dot{M}_\mathrm{Edd}\rangle$ & $\langle\Phi_{\rm BH}/(\dot{M}c)^{1/2}r_{\rm g}\rangle$ & $\langle P_{J}(100)\rangle$ [erg s$^{-1}$] & $\epsilon_J$ & $\langle P_{J,\rm{rad}}(100)\rangle$ [erg s$^{-1}$] & $\epsilon_{J,\rm{rad}}$ \\ 
\hline 
\texttt{H10}   &  0.9375 & Turb. Cascade & $3.6\times10^{-6}$ & 55  & $6.6\times10^{42}$ & 0.5 & $8.8\times10^{42}$& 0.7\\
\texttt{R17}   &  0.9375 & Mag. Reconnection  &  $2.3\times10^{-6}$ & 63 & $1.3\times10^{43}$& 1.6 & $1.4\times10^{43}$& 1.6 \\
\end{tabular}
\caption{
Time-averaged quantities for both simulations. From left to right, the quantities presented are the model name, the spin and heating prescription used, the average mass accretion rate through the black hole horizon measured in Eddington units, the magnetic flux threading the horizon measured in natural units, the mechanical jet power $P_{J}$ measured at a radius of $100\,r_{\rm g}$ (Eq.~\ref{eq::jetpower}), the corresponding jet efficiency $\epsilon_J =  \langle P_{J}\rangle/\langle\dot{M}c^2\rangle$, the jet power including radiation $P_{J,\rm{rad}}$ (Eq.~\ref{eq::jetpowerrad}) and the corresponding efficiency, both measured at the same radius.
}
\label{tab::summary}
\end{table*}

\subsection{Accretion Flow Properties}
\label{sec::properties}

In Figs.~\ref{fig::sym1} and ~\ref{fig::sym2}, we show quantities averaged in azimuth and time over the time period $t=11,000-16,000\,t_{\rm g}$ after rescaling the density, internal energy, and magnetic field to match the 230 GHz flux density measured by the EHT. 

Fig.~\ref{fig::sym1} shows properties related to the thermodynamics of the accretion flow:  the electron heating fraction $\delta_\mathrm{e}$, the gas temperature $T_{\rm gas}$, the electron temperature $T_\mathrm{e}$, the ion temperature $T_\mathrm{i}$, and the temperature ratio $T_\mathrm{e}/T_\mathrm{i}$. Fig.~\ref{fig::sym2} displays the mass density $\rho$, the bulk Lorentz factor $\gamma=u^0/\sqrt{-g^{00}}$, the magnetization $\sigma_\mathrm{i}$, the ratio of ion thermal pressure to magnetic pressure $\beta_\mathrm{i}$, and the ratio of radiation pressure to gas pressure in the fluid frame $\beta_R = \hat{E}/3 p$. 

In each profile in Figs.~\ref{fig::sym1} and ~\ref{fig::sym2}, the solid white contour shows the $\sigma_{\rm i}=1$ surface, while the dotted black contour shows the surface where the Bernoulli number $\mathrm{Be}=0.05$. Expressing $T^{\mu}_{\nu}$ in Boyer-Lindquist coordinates, the Bernoulli number is \citep{Narayan2012, Sadowski2013}
\begin{equation}
\label{eq::bernoulli}
\mathrm{Be} = -\frac{T^{t}_{\;\;t}+R^{t}_{\;\;t}}{\rho u^t} - 1. 
\end{equation}
For a cold unmagnetized fluid, $\mathrm{Be}=0.05$ corresponds to a flow velocity of ${\approx}0.3c$ at infinity. 

From the leftmost panels of Fig.~\ref{fig::sym1}, the different asymptotic behaviors of the two heating prescriptions introduced in Sec.~\ref{sec::heating} are evident in the average values of $\delta_{\rm e}$ in the jet region. In model \texttt{H10}, $\delta_{\rm e} \approx 1$ everywhere inside the jet region defined by the $\mathrm{Be}=0.05$ contour. Outside the highly magnetized funnel, $\delta_e$ drops to nearly zero in the outer regions of the disc, $r\gtrsim40\,r_{\rm g}$. In contrast, in model \texttt{R17}, $\delta_{\rm e}$  reaches its limit of equipartition of thermal energy ($\delta_{\rm e}=0.5$) inside the $\sigma_{\rm i}=1$ contour. In the less magnetized  disc outside $r\gtrsim15\,r_{\rm g}$, $\delta_{\rm e}$ falls to a small but nonzero value $\delta_{\rm e}\approx0.2$.

While the temperature distribution of the combined gas is similar in the two models (the second column of Fig.~\ref{fig::sym1}), the different heating prescriptions result in different electron temperatures and temperature ratios in the inner disc and jet. In model \texttt{H10}, the deposit of nearly all thermal energy into electrons inside the jet results in high electron temperatures $T_{\rm e}\sim10^{11}$ K near the black hole that climb to $10^{12}$ K in the jet around 50 $r_{\rm g}$. While the temperature ratio $T_{\rm e}/T_{\rm i}$ is less than unity in the regions closest to the black hole, it rises above unity by $20 \, r_{\rm g}$ along the jet.  

In contrast, in the magnetic reconnection heated model \texttt{R17}, $T_{\rm e}/T_{\rm i} < 1$ everywhere. In the jet around 30 $r_{\rm g}$ from the black hole, $T_{\rm e}/T_{\rm i} \approx 0.3$, and it increases with radius, reaching 0.75 around 1000 $r_{\rm g}$. In the disc, while the value of $\delta_{\rm e}$ is higher than in the turbulent cascade heating simulation \texttt{H10}, the disc temperature ratio is not substantially different, taking an average value of ${\sim}0.1$ around $30 \, r_{\rm g}$. This was not the case in the \sgraa\ simulations presented in \citet{Chael18}, where the turbulent cascade heated simulations had lower electron temperatures in the disc than the simulations heated by magnetic reconnection. The similarity of the outer disc electron temperatures in the present models may arise from the increased importance of Coulomb coupling in the denser regions of these higher accretion rate simulations \citep{RyanM87}. 

Fig.~\ref{fig::sym1} shows that, as in the simulations of \sgraa\ \citep{Chael17}, the choice of electron heating prescription has a noticeable effect on the electron-ion thermodynamics of the system. Unlike in the low-magnetic-flux simulations considered in that work, however, the two MAD simulations considered here have notably different gas and radiation kinematics as well, arising from the choice of heating prescription, even though both models produce thick (scale height-to-radius ratio $H/R \approx 0.41$ at $10\,r_{\rm g}$), highly magnetized discs.

Table~\ref{tab::summary} summarizes important time-averaged quantities from our simulations, and Fig.~\ref{fig::acc_mad} shows the accretion rate, magnetic flux through the horizon, and 230 GHz flux density as a function of time. Both simulations have a low accretion rate ${\sim}10^{-6} \dot{M}_{\rm Edd}$, and both simulations reach the MAD state, with the magnetic flux threading the black hole $> 50\,\sqrt{\dot{M}c}\,r_{\rm g}$, though \texttt{R17} is slightly more magnetized. In both simulations, while the averaged value of the accretion rate is stable, there are larger excursions with time than in the low-magnetic-flux simulations of \citet{Chael18} due to the interaction of the accretion flow and the magnetic flux. This is particularly apparent in \texttt{R17}, which is more magnetized. 

The high jet electron temperatures arising from the turbulent cascade heated model in  simulation \texttt{H10} produce a much more intense radiation field from synchrotron and inverse Compton scattering in the jet region than in \texttt{R17}. This is easily seen in the last column of Fig.~\ref{fig::sym2}, which shows the ratio of the radiation pressure to gas pressure as measured in \texttt{KORAL}. While this quantity is ${\lsim}1$ almost everywhere in the \texttt{R17} simulation, it approaches values ${\gtrsim} 100$ in the jet in model \texttt{H10}. 

While \texttt{H10} produces more powerful synchrotron and inverse Compton radiation than \texttt{R17}, the optical depth to Compton scattering in both discs is low, with $\tau_{\rm IC}\sim10^{-2}$. For the less magnetized models of \citet{RyanM87}, in contrast, the higher densities and temperatures needed to match the M87 spectrum with weaker magnetic fields made inverse Compton scattering more efficient, with $\tau_{\rm IC}\sim0.1-1$.

\begin{figure}
\centering
\includegraphics*[width=0.495\textwidth]{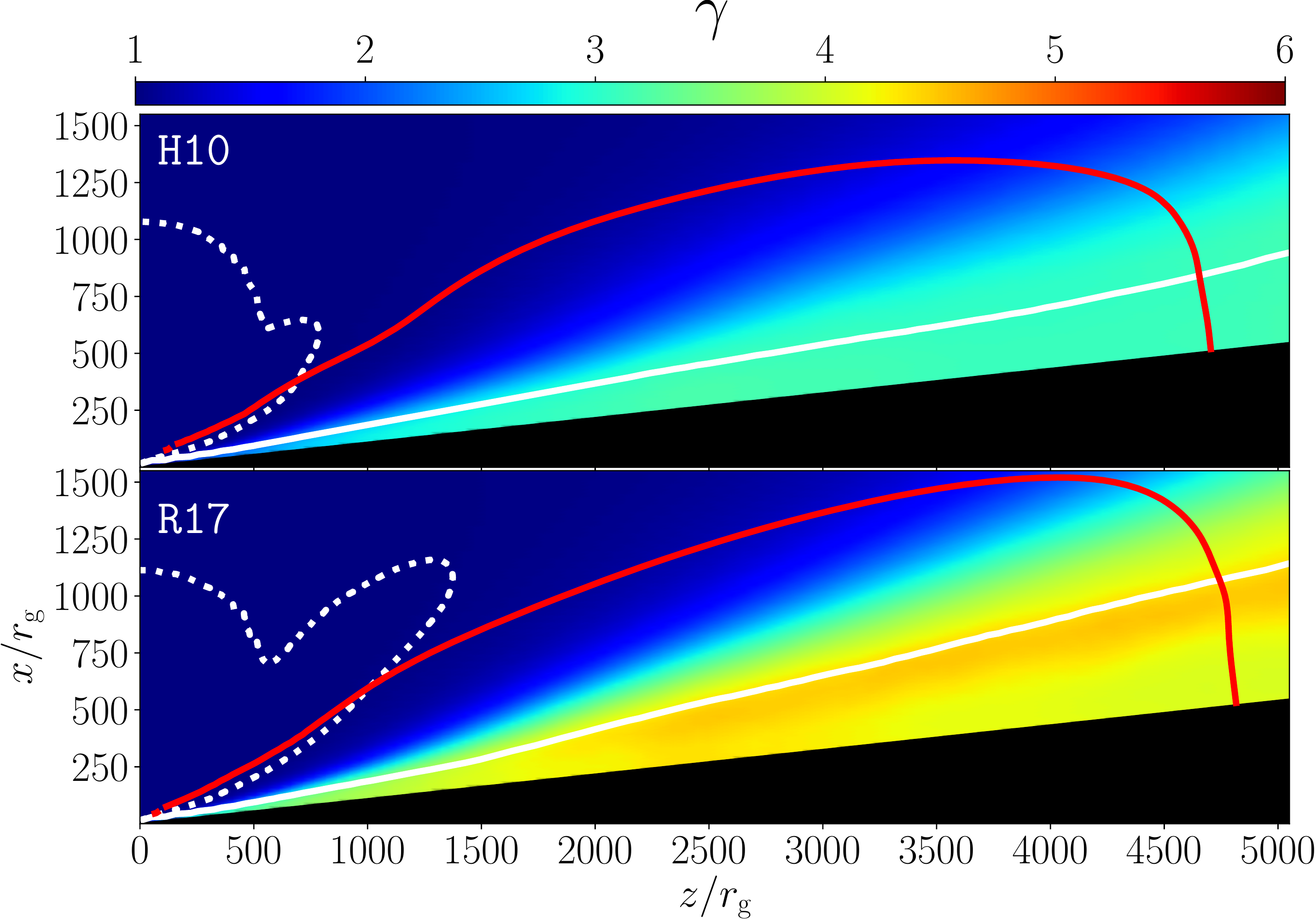}
\caption{
Large-scale jet Lorentz factors $\gamma$ from the simulations \texttt{H10} (top) and \texttt{R17} (bottom). The solid white contour shows the surface where $\sigma_{\rm i}$=1, and the dashed white contour shows where the Bernoulli parameter ${\rm Be} = 0.05$. The red contour indicates the extend of the region of inflow/outflow equilibrium,  where $t_{\rm acc} = 5000 \, t_{\rm g}$ (Eq.~\ref{eq::inflow}). The black region along the jet axis indicates the two nearest cells to the polar axis, which are removed from all analysis due to their boundary conditions (Section~\ref{sec::setup}). 
}
\label{fig::sym_lorentz}
\end{figure}

The conversion of much of model \texttt{H10}'s energy and momentum to radiation in the inner jet has a significant impact on its mechanical properties. While both simulations launch relativistic jets. \texttt{H10}'s jet is weaker, with Lorentz factor $\gamma \approx 1.5$ at $50 \, r_{\rm g}$ compared to $\gamma \approx 2$ for \texttt{R17} (Fig.~\ref{fig::symc2}, second column). Fig.~\ref{fig::sym_lorentz} shows the Lorentz factors of the jets at large scales. By the jet equilibrium radius around $4700\,r_{\rm g}$, \texttt{H10} reaches a Lorentz factor of $\gamma \approx3$, while \texttt{R17} reaches $\gamma \approx 5$. The conversion of fluid and magnetic energy into radiation also likely accounts for \texttt{H10}'s smaller value of horizon flux $\langle\Phi_{\rm BH} \rangle$ and correspondingly larger mass accretion rate $\dot{M}$ (Table~\ref{tab::summary}). 


At large radii, $r>1000\,r_{\rm g}$, our simulations resolve the jet with ${\approx} 6$ cells in polar angle out to the $\sigma_{\rm i} = 1$ contour (white lines in Fig.~\ref{fig::sym_lorentz}). Higher resolution simulations will be necessary to test whether the observed jet opening angle in simulated images from these simulations (Section~\ref{sec::images}) is affected by the simulation resolution. Furthermore, our choice of reflective boundary conditions along the polar axis (Section~\ref{sec::setup}) may enlarge the jet width; tests of similar MAD jets in Cartesian coordinates \citep[e.g.][]{Bhac} or using a misaligned grid \citep[e.g.][]{Liska2018} are necessary to assess the dependence of apparent jet width on the numerical grid. 

We measure the thermal, magnetic, and jet mechanical power in both simulations as a function of radius using the definition \citep{Tchekhovskoy11,RyanM87}
\begin{equation}
 \label{eq::jetpower}
 P_J = -\int \left( T^{r}_{\;\;t}  + \rho u^r\right)\,\sqrt{-g}\,\mathrm{d}\theta\mathrm{d}\phi,
\end{equation}
where the integral is at a fixed $r$ is over the jet cap, which we define by the criterion $\mathrm{Be}>0.05$ \citep{Narayan2012,Sadowski2013}. The time-averaged jet power measured by Eq.~\ref{eq::jetpower} is roughly constant with radius from around $r=10\,r_{\rm g}$ out to $r=1000\,r_{\rm g}$. We measure the average jet powers at $100\,r_{\rm g}$ from the averaged data to be $6.6\times10^{42}\,\text{erg s}^{-1}$ for model \texttt{H10} and $1.3\times10^{43}\,\text{erg s}^{-1}$ for \texttt{R17} (Table~\ref{tab::summary}).

While the jet powers obtained from the two simulations agree to within a factor of two, the value obtained for model \texttt{R17} is more consistent with the measured values for M87 of ${\sim}10^{43} - 10^{44}\text{erg s}^{-1}$ \citep{Reynolds96_2,Stawarz06}.  Comparing the jet power to the accretion rate gives a jet efficiency $\epsilon_J = P_J/\dot{M}c^2$ of 1.6 and 0.5 for \texttt{R17} and \texttt{H10}, respectively, indicating that spin energy is being extracted from the black hole. This is especially true in model \texttt{R17}, which has greater than 100 per cent efficiency \citep{Tchekhovskoy11}. 

Because much of \texttt{H10}'s energy and momentum is converted to radiation in the jet, it has a correspondingly lower mechanical jet power. Including radiation in the jet power measurement, we define
\begin{equation}
 \label{eq::jetpowerrad}
 P_{J,\rm{rad}} = -\int \left( T^{r}_{\;\;t} + R^{r}_{\;\;t} -\rho u^r\right)\,\sqrt{-g}\,\mathrm{d}\theta\mathrm{d}\phi,
\end{equation}
This increases the measured jet powers to $P_{J,\rm{rad}}=8.8\times10^{42}$ erg s$^{-1}$ for \texttt{H10} and $P_{J,\rm{rad}}=1.4\times10^{43}$ erg s$^{-1}$ for \texttt{R17}, and increases the jet efficiencies in the two models to 0.7 and 1.6, respectively.

\begin{figure*}
\centering
\includegraphics[width=0.99\textwidth]{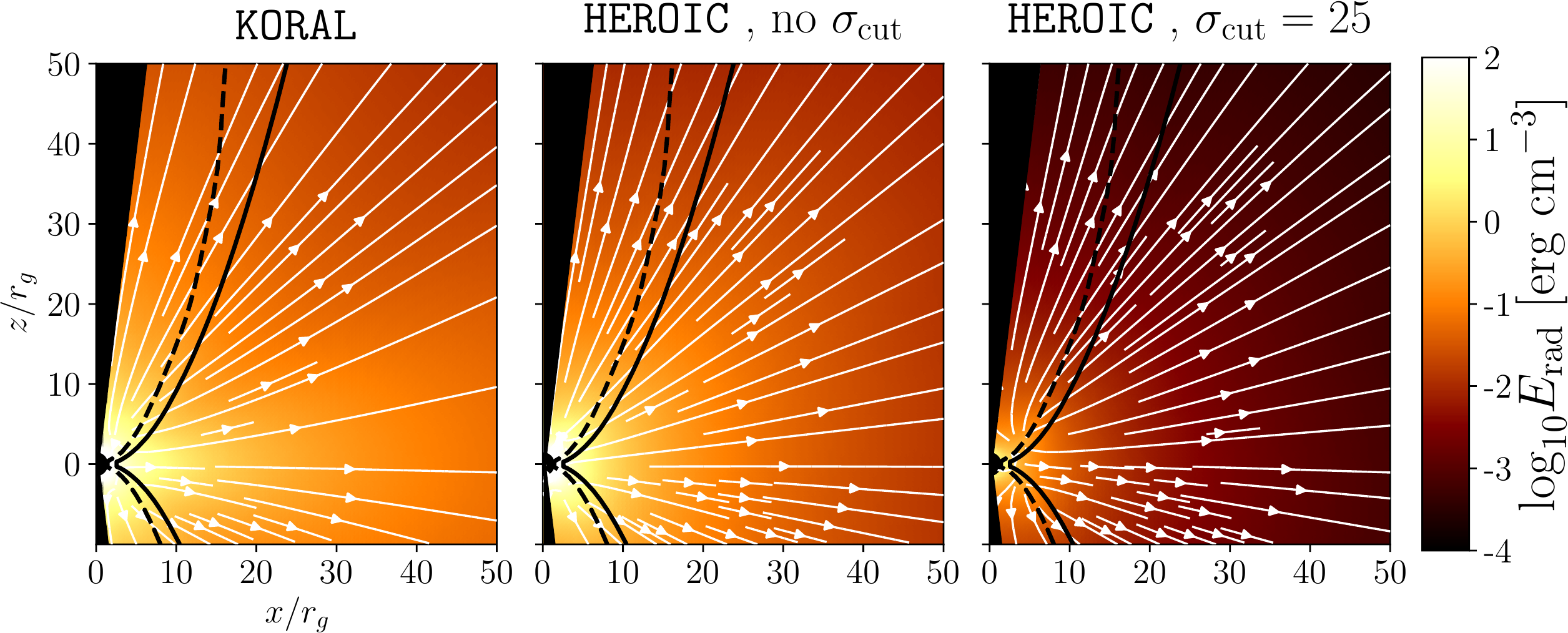}
\caption{
(Left) Lab frame radiation energy density $E_{\rm rad}$ and radiation flux streamlines from time- and azimuth- averaged data from simulation \texttt{H10}. The solid contour indicates the $\sigma_{\rm i}=1$ surface, and the dashed contour indicates $\sigma_{\rm i} = \sigma_{\rm cut} = 25$. (Center) the same quantities computed in the postprocessing code \texttt{HEROIC}, with no cut on $\sigma_{\rm i}$ imposed.  (Left) the radiation  energy density and flux streamlines from \texttt{HEROIC} after zeroing emissivities from the regions $\sigma_{\rm i} < \sigma_{\rm cut}$. The \texttt{KORAL} and \texttt{HEROIC} results are in good agreement when no cut on radiation from $\sigma_{\rm i}> 25$ regions is included; both codes produce approximately radial radiation streamlines and extremely high luminosities originating from near the black hole. When the $\sigma_{\rm i}$ cut is imposed in the \texttt{HEROIC} post-processing, the average energy density in radiation  drops by more than a factor of 10.
}
\label{fig::sym_radflux}
\end{figure*}

Much of the intense radiation in \texttt{H10} is produced from regions near the $\sigma_{\rm i, max}=100$ ceiling. Because mass is constantly being injected in these regions, energy and momentum is added to the simulation and is then efficiently converted into radiation. As discussed in Section~\ref{sec::radiation}, we do not trust the radiation produced in this region, and we do not include these regions in our post-processing computation of spectra and images in the following sections. Fig.~\ref{fig::sym_radflux} shows a comparison of the average radiation field from the \texttt{KORAL} output for model \texttt{R17} with the average radiation fields computed from \texttt{HEROIC} both with no $\sigma_{\rm cut}$ imposed and using $\sigma_{\rm cut} = 25$. With no $\sigma_{\rm cut}$, both codes produce consistent results, with radial radiation streamlines and extremely high energy radiation energy densities. However, when we emissivities from regions with $\sigma_{\rm i} >25$ are zeroed out in \texttt{HEROIC} (as is done to produce the spectra and images in Sections~\ref{sec::spectra} and~\ref{sec::images}) the energy density of the radiation field everywhere drops by a  factor of ${\approx}50$.

In the (optically thin) GRRMHD simulation itself, because the frequency-averaged radiation field produced in $\sigma>25$ regions spreads through the simulation volume at nearly the speed of light, it is difficult to extract meaningful radiation quantities that are unaffected by the density floors.
For these reasons, when interpreting the jet power and other quantities from these simulations, it is important emphasize again that the results may be strongly dependent on the specific choice of density floors. More work is needed to better understand the impact of radiation from high $\sigma$ regions on the global structure of optically thin accretion flows; future two-temperature MAD simulations should consider not including radiation from $\sigma>\sigma_{\rm cut}$ regions during the simulation run.

\subsection{Spectra}
\label{sec::spectra}
\begin{figure*}
\centering
\includegraphics*[width=0.99\textwidth]{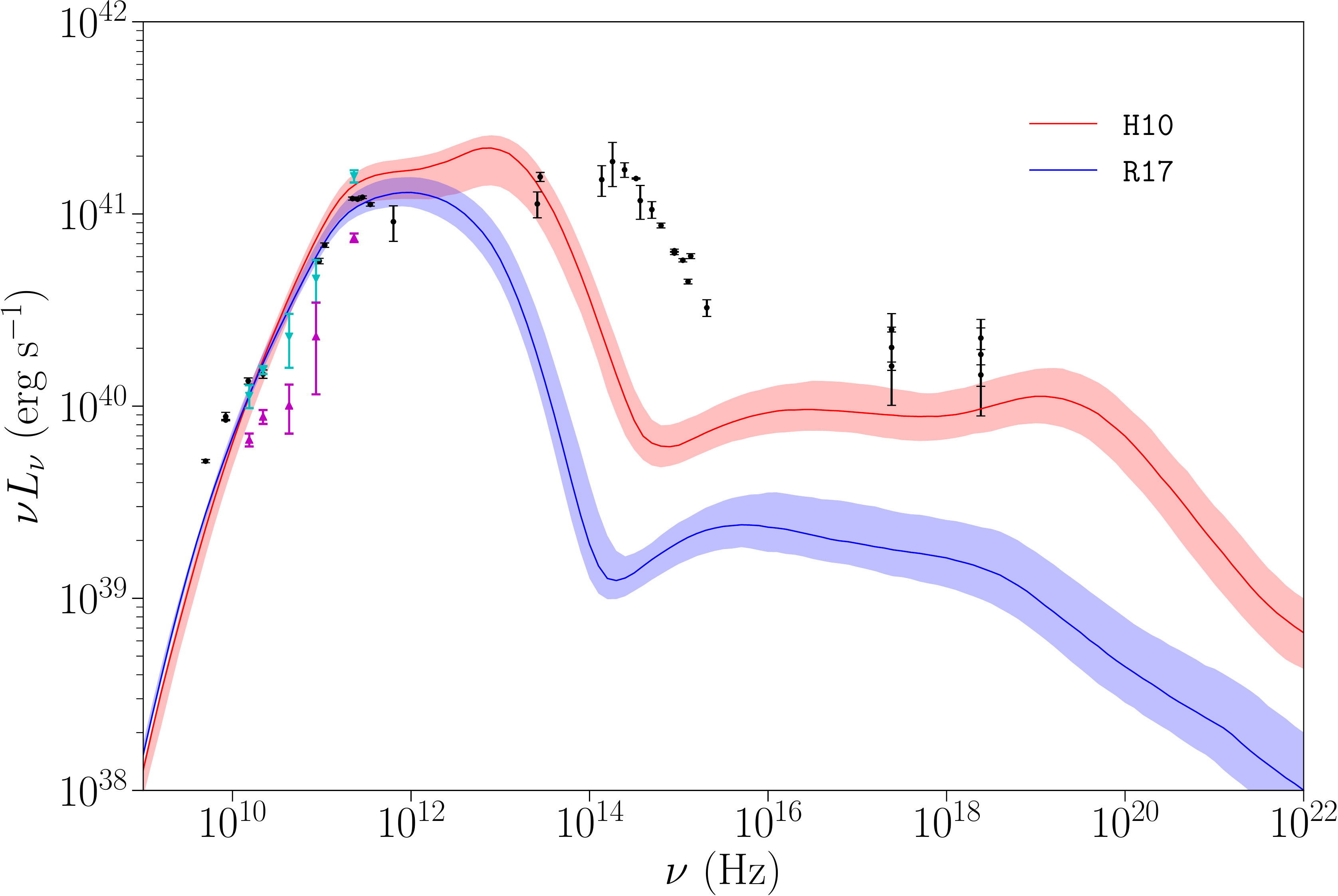}
\caption{
SEDs for the two models calculated with \texttt{HEROIC} for an observer at $17^\circ$ inclination \citep{Walker2018} measured up from the simulation south pole. Spectra were computed from 3D simulation snapshots every $10 \, t_{\rm g}$ over the $5000 \,t_{\rm g}$ period from $t=11,000\,t_{\rm g}$ to $t=16,000\,t_{\rm g}$ after rescaling the density to approximately match the 0.98~Jy flux density of compact emission at 230 GHz \citep{Doeleman12,Akiyama15}. The solid curve shows the median spectrum for each model, and the shaded region shows the nominal $1\sigma$ time-variability if we assume that the variability distribution is Gaussian at each frequency. Data points are taken from Table 1 of \citet{Prieto16} (black) and Table~\ref{tab:M87_compact_flux} of this work. Measurements of the total flux density at radio wavelengths from Table~\ref{tab:M87_compact_flux} are displayed in cyan, while measurements of the compact flux density of the core are displayed in magenta.  
}
\label{fig::spectra}

\captionof{table}{The total and compact radio spectrum of M87 from recent VLBI observations.}
\label{tab:M87_compact_flux}
\begin{tabular}{cccl}
\hline
Frequency & Total Flux Density & Core Flux Density & Data Used\\
$\text{[GHz]}$ & $\text{[Jy]}$ & $\text{[Jy]}$ & \\
\hline
15.4  & $2.2 \pm 0.3$   & $1.3 \pm 0.1$  & 19 MOJAVE observations from 2001-2011 \citep{Lister2018}.\\ 
22    & $2.1 \pm 0.1$   & $1.2 \pm 0.1$  & 10 KaVA \& 3 VLBA (24 GHz) observations from 2013-2014 \citep{Hada2017}.\\
43.1  & $1.6 \pm 0.4$   & $0.7 \pm 0.2$  & 50 VLBA observations from 1999-2016 \citep[Table~3 in][]{Walker2018}.\\ 
86.3  & $1.1 \pm 0.5$   & $0.8 \pm 0.4$ &  5 GMVA observations analyzed by \citet{KimM87}.\\
230.0 & $2.05 \pm 0.15$ & $0.98 \pm 0.05$ & 2 EHT observations: \citet{Doeleman12} \& \citet{Akiyama15}.\\
\hline
\end{tabular}
\end{figure*}

Fig.~\ref{fig::spectra} shows the spectral energy distributions (SEDs) for both of our simulations. These were obtained from postprocessing with \texttt{HEROIC} over the full $5000 \, t_{\rm g}$ duration of the simulation from $11,000$-$16,000\,t_{\rm  g}$, $1000\,t_{\rm  g}$ after rescaling the density to approximately match the compact 230 GHz flux measured by \citet{Doeleman12} and \citet{Akiyama15}. Synchrotron, free-free, and inverse Compton emission are all included in the \texttt{HEROIC} calculations, although we found that free-free emission does not contribute significantly even in the X-ray. These spectra do not include radiation produced from regions with $\sigma_{\rm i}>25$, and consequently the total luminosity from the postprocessing spectra is less than that produced in the simulation bolometric $R^{\mu}_{\;\;\nu}$. In computing spectra, \texttt{HEROIC} used data from the simulation out to $1000 \, r_{\rm g}$, so diffuse emission from a good fraction of the jet is included, but the outermost regions of the jet are ignored. 

Comparing simulated SEDs with observations of M87 requires some care. Specifically, because the total radio flux density along the extended jet of M87 is comparable to that of the significantly brighter but compact region near the black hole (the ``core''), a meaningful comparison with observations requires excising jet contributions that are outside our simulated domain. \citet{Prieto16} have compiled measurements (their Table~1) from radio to X-ray of the M87 spectrum in a quiescent state, using only measurements that achieve at least $0.4''$ resolution in order to securely exclude emission from the brightest jet knot, HST-1. They also compiled measurements of the total flux density of the most compact component identified by VLBI observations (their Table~4). However, these latter measurements have some notable limitations. For example, at 86\,GHz, \citet{Prieto16} include the value $S_{\rm L} = 0.16 \pm 0.07\,{\rm Jy}$ reported by \citet{Lee2008}, corresponding to the measured flux density on the longest baseline for observations with the Coordinated (global) Millimeter VLBI Array (CMVA/GMVA) 
This approach is problematic because the core is resolved at 86 GHz \citep{KimM87} and because interference among compact components can significantly affect the correlated flux density on a single baseline. At 22 GHz,they include the compact flux density value 0.35~Jy, reported by \citet{Junor1995}. However, during the epochs reported by \citet{Junor1995}, the total flux density of M87, including extended structure, was only ${\sim}1.1\,{\rm Jy}$, which is significantly lower than the values measured more recently with the Very Long Baseline Array (VLBA) and the KVN/VERA Array (KaVA) \citep{Hada2017}. It is also lower than the values measured at 15\,GHz and 43\,GHz since 2000 \citep{Lister2018, Walker2018}.

Because we normalize our results to have a total flux density at 230\,GHz that matches measurements taken in 2009 and 2012, we provide updated estimates of the total and compact  flux density of M87 in Table~\ref{tab:M87_compact_flux}. To estimate the flux density of the compact component, we measure the peak flux density of a beam convolved image from VLBI observations at a given frequency.
This procedure gives a direct comparison between our simulated images and reconstructed images from VLBI. We note that this total flux density measured with VLBI may still contain significant contributions from outside our simulation domain, especially at centimeter wavelengths.%


SEDs from both our models are largely consistent with the radio spectrum data up to the synchrotron peak at 230\,GHz; this flat spectrum is also produced by analytic models of relativistic jets \citep{BK79, Falcke95}. The model spectra underpredict the total measured flux at frequencies $<10^{10}$ Hz; at these low frequencies, the jet on scales larger than $1000 \, r_{\rm g} \approx 3600 \mu$as likely makes substantial contributions to the total emission. 

Neither SED matches the measured flux at infrared through ultraviolet frequencies, although the hot jet electrons in the \texttt{H10} model do extend the thermal synchrotron spectrum to the ${\approx}3\times10^{13}~{\rm Hz}$ measurements by \citet{Perlman01} and \citet{Whysong04}. The observed emission from the near infrared to ultraviolet may be explained by the addition of a high-energy non-thermal electron population in the disc \citep{Broderick09} or the jet \citep{Dexter12,Prieto16}.

The hotter electrons in the \texttt{H10} simulation produce more inverse Compton power at X-ray frequencies. Both models produce flat SEDs from Comptonization at X-ray frequencies, which roughly match the slope of the \emph{Chandra} measurements of the core of M87 \citep{DiMatteo2003}, but underpredict the total flux density in the X-ray. However, the shape of the spectrum at frequencies $>10^{12}$ Hz depends strongly on the choice of $\sigma_{\rm cut}$ (see Section~\ref{sec::sigcut}). It is possible that a two-temperature MAD simulation that can be trusted up to higher values of $\sigma_{\rm i}$ could fit all of the spectral data up to the X-ray. Neither model's SED extends significantly into the $\gamma$-ray, where the observed emission may be dominated by the jet knots, such as HST-1 \citep{Abramowski2012}. 

\subsection{Variability}
\label{sec::variability}

\begin{figure}
\centering
\includegraphics*[width=0.5\textwidth]{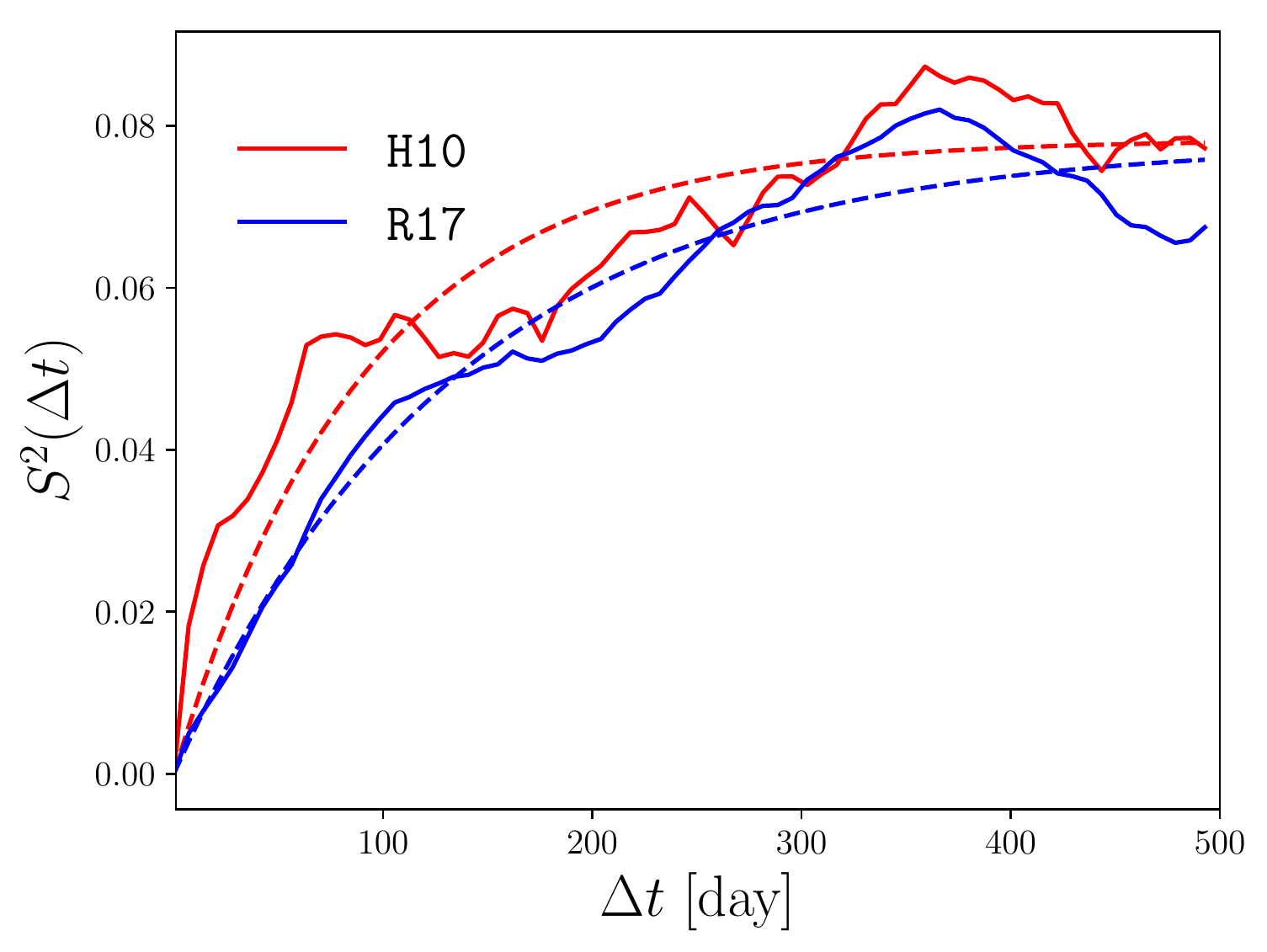}
\caption{Structure functions of the 230 GHz light curves produced by the two simulations. The structure function computed from the data with  Eq.~\eqref{eq::structfunc} is a solid line for each simulation and the least-squares fit to  the damped random walk form (Eq.~\ref{eq::structfuncdrw}) is displayed with a dashed line. Model \texttt{H10} has a time-scale $\tau = 01$~days, while \texttt{R17} has $\tau = 135$ days. \citet{Bower2015b} measure $\tau=45^{+61}_{-24}~{\rm days}$ using more than 10 years of data taken with the SMA.
}
\label{fig::structfunc}
\end{figure}

The 230 GHz lightcurves from our high-resolution \texttt{grtrans} images are presented in the bottom panel of Fig.~\ref{fig::acc_mad}. Both simulations show variability on time-scales of days to years, and the normalized root-mean-square variability of both lightcurves is ${\approx} 15-20$ per cent relative to the mean. 

\citet{Bower2015b} modeled the 230 GHz light curve of M87 observed over more than 10 years with the Submillimeter Array (SMA) as a damped random walk process. They define the structure function $S^2(\Delta t)$ of the lightcurve $s(t)$ as 
\begin{equation}
 \label{eq::structfunc}
 S^2(\Delta t) = \frac{1}{N}\sum\left(s(t) - s(t + \Delta t)\right)^2,
\end{equation}
where $N$ is the number of data points in the sum. For a damped random walk, this structure function takes the form
\begin{equation}
\label{eq::structfuncdrw}
 S^2_{\rm DRW}(\Delta t) = S^2_\infty\left(1-\text{e}^{-\Delta t/\tau}\right),
\end{equation}
where $S^2_\infty$ is the power on long time-scales and $\tau$ is the correlation or damping time of the random walk. 

For real data that is noisy and irregularly spaced, the structure function defined in Eq.~\eqref{eq::structfunc} can have spurious artefacts. Therefore, \citet{Bower2015b} did not measure the structure function  directly but instead used a Bayesian approach to model the irregularly sampled lightcurve and extract the variability time-scale (see \citealt{Dexter14} for details). For the SMA observations of M87, they measured a damped random walk time-scale $\tau=45^{+61}_{-24}~{\rm days} = 127^{+173}_{-68}\,t_{\rm g}$. 

Here, since we have regularly sampled, noise-free lightcurves from our simulations over ${\sim}5$ years, to measure the correlation time-scale in our 230 GHz lightcurves we simply compute the structure function and perform  a least squares fit to obtain $S^2_\infty$ and $\tau$. We compute the structure function for time-scales $\Delta t$ ranging from our sampling cadence $10\,t_{\rm g} = 3.5$ days up to 500 days. We first remove a linear slope from both lightcurves before computing the structure functions. For \texttt{R17} this slope is negligible, but as discussed in Sec.~\ref{sec::properties}, simulation \texttt{H10} shows a slight but steady increase of 230 GHz flux density over the interval considered; the slope we subtract before performing the structure function analysis is ${\approx} 0.2 \, {\rm Jy} / 5000 \, t_{\rm g}$.  

Our structure functions and the best-fitting damped random walk models (Eq.~\ref{eq::structfuncdrw}) are presented in Fig.~\ref{fig::structfunc}. The structure functions of both models show the characteristic form of a damped random walk, with similar time-scales.  For \texttt{H10}, we measure a correlation time-scale $\tau \approx 91$~days, while for \texttt{R17} we measure $\tau \approx 135$~days. 

Both models have variability time-scales ${\sim}2$ times longer than the measurement of \citet{Bower2015b}. However, these results should be viewed with some caution, as our lightcurves cover only $4.8$ yr and thus may not be long enough to robustly determine a long-term variability timescale.  Furthermore, the grid resolution of our simulation sets a fundamental limit to the scale of our turbulent eddies, and hence, the fast variability timescale. In the Appendix, we show that the variability amplitude is not consistent between simulation \texttt{R17} and an identical simulation run at lower resolution. It is possible an even higher-resolution simulation with the same parameters would show faster variability with a lower amplitude. 

Furthermore, our lightcurves were produced with the ``fast light'' approximation, ignoring the evolution of the fluid as photons propagate through the accretion flow and jet. Properly accounting for the fluid evolution during the radiative transfer may significantly change the characteristics of the produced variability.

\subsection{Images}
\label{sec::images}

\begin{figure*}
\centering
\includegraphics*[width=0.99\textwidth]{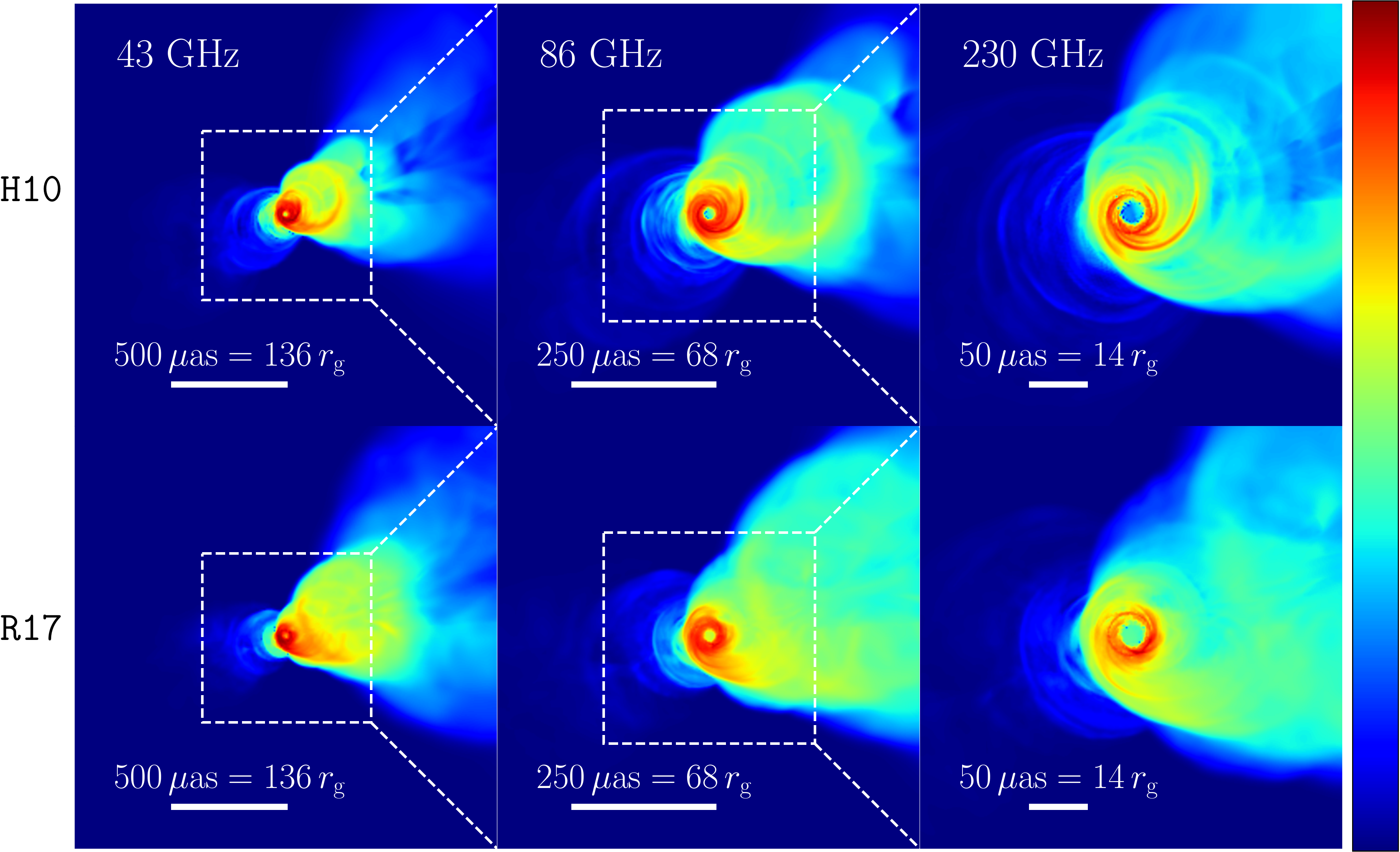}
\caption{Log scale images of simulation snapshots of the two models at 43 GHz (left), 86 GHz (middle) and 230 GHz (right). Snapshots were observed at an inclination angle of $17^\circ$ up from the simulation south pole and rotated $108^\circ$ counterclockwise to match the observed jet orientation. The intensity scale is different at each frequency, but for each frequency the scale displays a dynamic range of $10^4$ and is the same for both the image from the \texttt{H10} simulation (top) and \texttt{R17} simulation (bottom). The image length scale changes with frequency; dotted boxes on the 43 and 86 GHz images show the fields-of-view of the 86 and 230 GHz images, respectively. The jet structure is qualitatively similar in the two simulations, with wide apparent opening angles which narrow with distance from the SMBH at lower frequencies. Images at all frequencies show a faint counterjet, and the black hole shadow is evident in both models even down to 43 GHz.
}
\label{fig::logimages}
\end{figure*}

\begin{figure}
\centering
\includegraphics[width=0.48\textwidth]{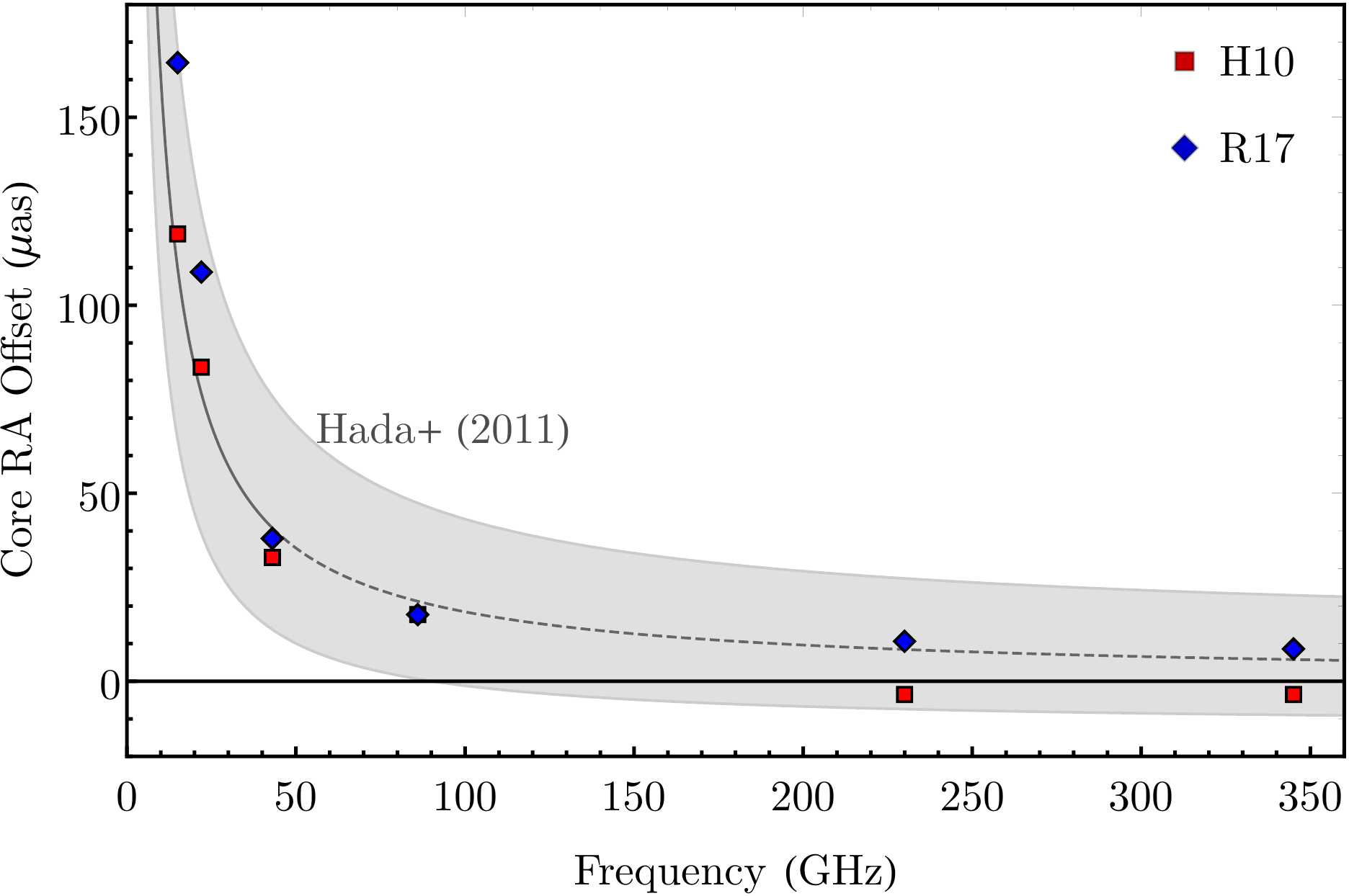}
\caption{Frequency dependent core-shift for \texttt{H10} and \texttt{R17} (see Section~\ref{sec:core_shift}). The gray line and shaded region show the best-fit model and $1\sigma$ confidence region from \citet{Hada2011} using measurements from 2 - 43~GHz (here, we re-reference the central model to a core-shift of zero as $\nu \rightarrow \infty$). All points shown are compatible with the VLBI measurements over this frequency range. We show the extrapolated model and our corresponding simulation estimates at higher frequencies, but the latter must be interpreted with caution because the images are no longer dominated by a bright, optically-thick core.
}
\label{fig::core_shift}
\end{figure}

\begin{figure*}
\centering
\includegraphics*[width=0.99\textwidth]{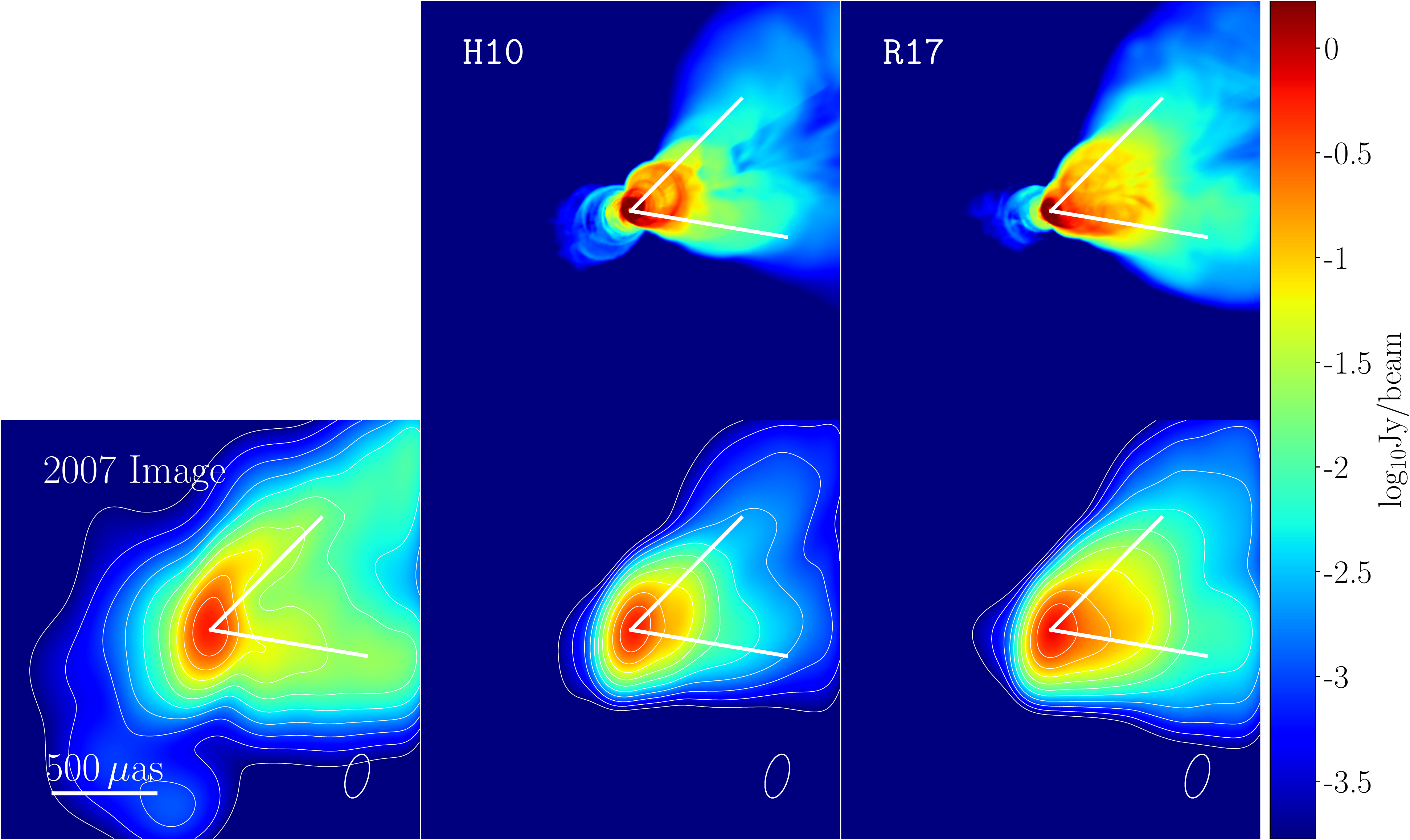}
\caption{Log scale 43 GHz images of snapshots from the two models overlaid with the measured $55^\circ$ apparent opening angle reported by \citet{Walker2018}. The leftmost panel shows an image reconstructed from 2007 VLBA data \citep{Walker2018} using the \texttt{eht-imaging} library \citep[Fig. 9 of][]{Chael18_Closure}. The image is convolved with a Gaussian beam half the size of the nominal beam reported in \citet{Walker2018}. The top row shows the high-resolution \texttt{grtrans} images of the two simulations at 43 GHz (zoomed out versions of the 43 GHz images in Fig.~\ref{fig::logimages}). The bottom row shows the simulation images convolved with the same beam as the 2007 reconstruction. The snapshot images were normalized to the total flux of the 2007 image. The MAD simulations in this work naturally produce wide opening angle jets at 43 GHz, but they show less counterjet emission than the 2007 image.
}
\label{fig::im43}
\end{figure*}

\begin{figure*}
\centering
\includegraphics*[width=0.99\textwidth]{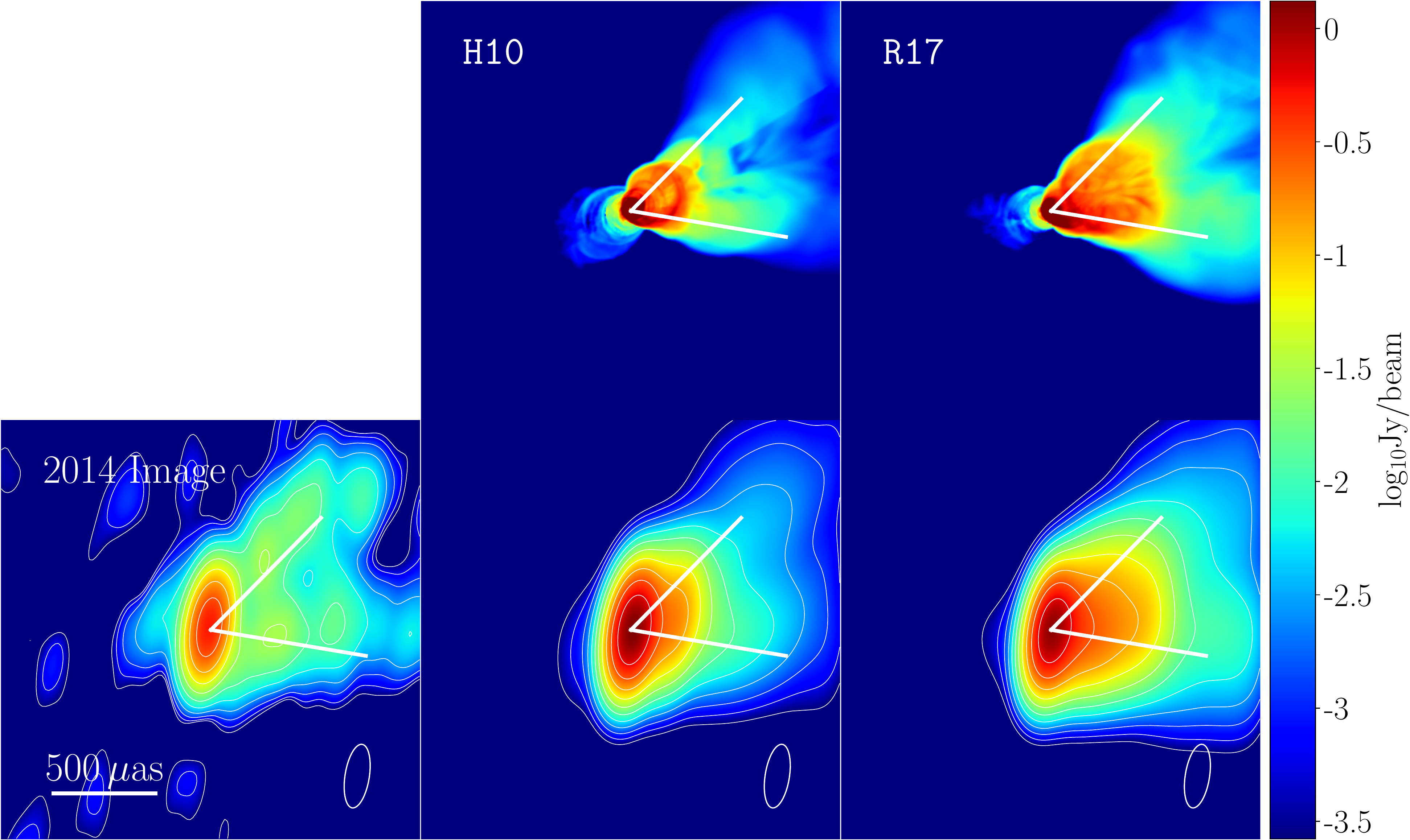}
\caption{Log scale images of 86 GHz snapshots of the two models overlaid with the measured $55^\circ$ apparent opening angle. The leftmost panel shows an image reconstruction using the CLEAN algorithm on 2014 data from the GMVA reported in \citet{KimM87}. The top row shows the high-resolution \texttt{grtrans} images of the two simulations at 86 GHz  (zoomed out versions of the 86 GHz images in Fig.~\ref{fig::logimages}). The bottom row shows the simulation images convolved with the Gaussian beam reported in \citet{KimM87}. The snapshot images were normalized to the total flux of the 2014 image. Both simulations produce wide opening angle jets at 86 GHz and noticeable counterjet emission, although the limb brightening in the blurred simulation images is somewhat less than in the 86 GHz CLEAN reconstruction.
}
\label{fig::im86}
\end{figure*}

\begin{figure*}
\centering
\includegraphics*[width=0.99\textwidth]{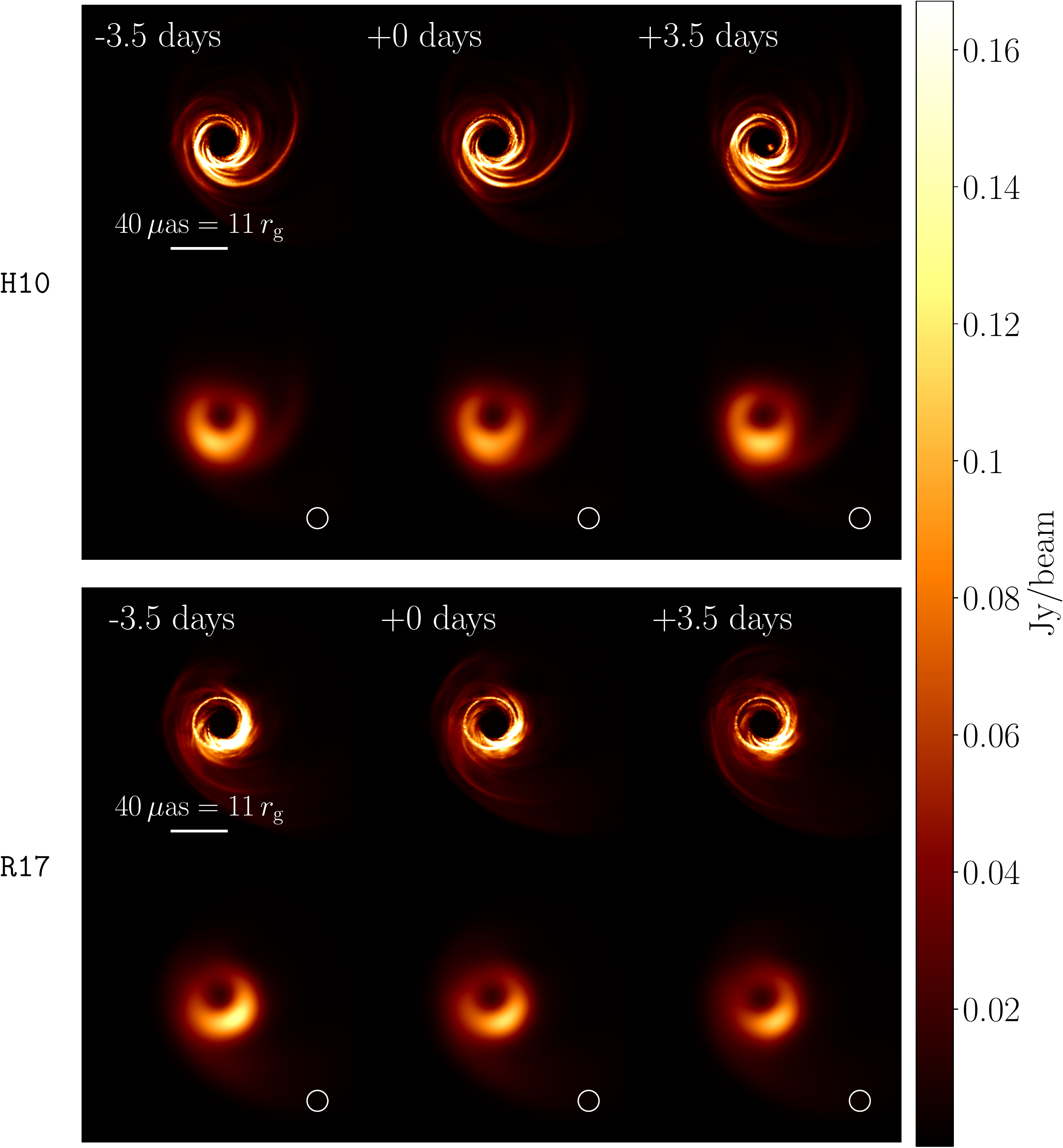}
\caption{
Linear scale images of sequential snapshots at 230~GHz from two models showing time evolution over $20 \, t_{\rm g}\approx 7\,{\rm days}$, the usual length of an EHT observing campaign. The top two rows show images from simulation \texttt{H10}, and the bottom two rows show images simulation \texttt{R17}. In each set of images, the top row shows high-resolution images from \texttt{grtrans}, and the bottom row shows the same images blurred with a circular Gaussian beam with a $15\,\mu$as FWHM, approximately representing the imaging resolution of the EHT. 
}
\label{fig::im230}
\end{figure*}

\begin{figure}
\centering
\includegraphics*[width=0.4\textwidth]{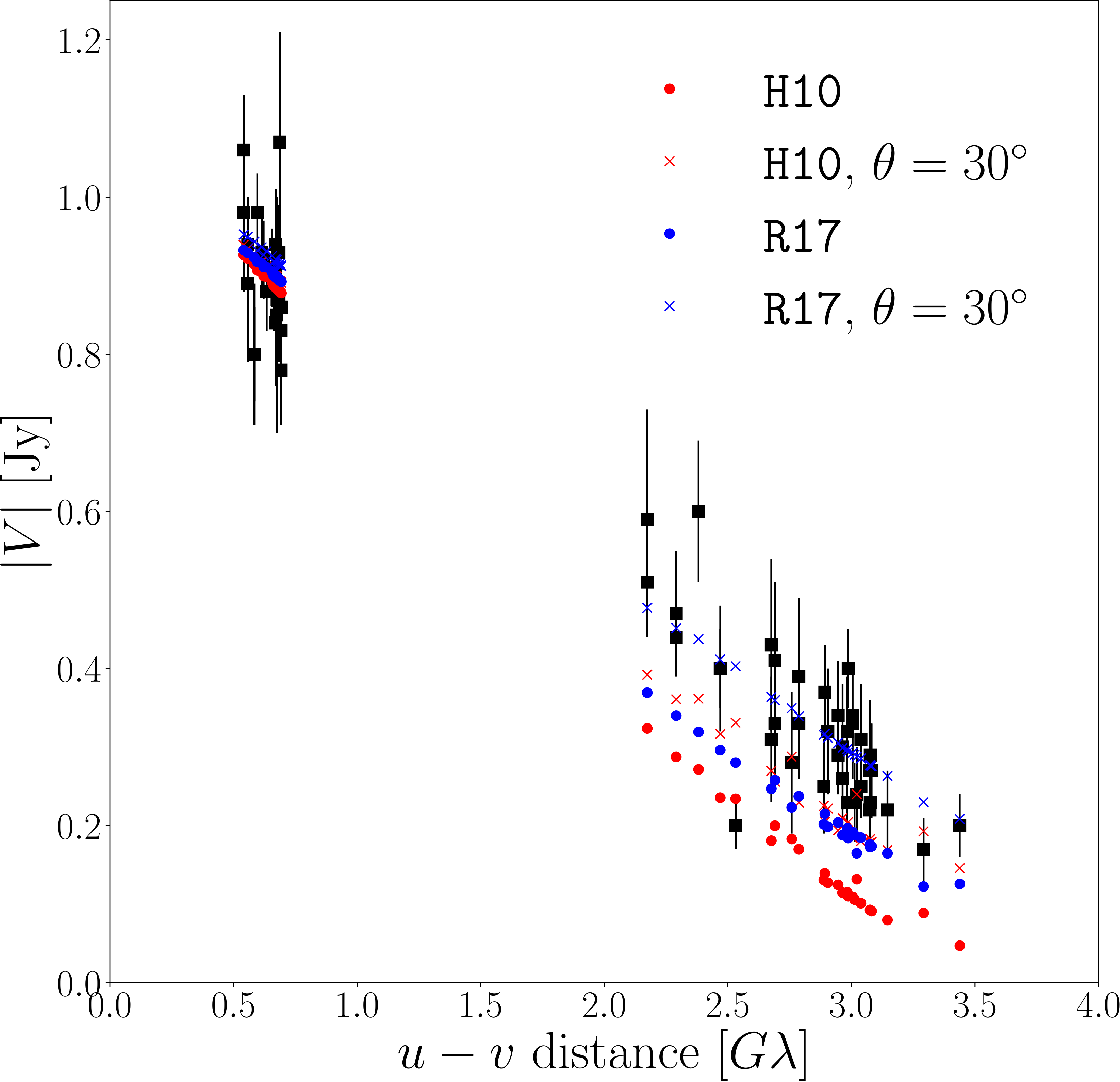}
\caption{Comparison of visibility amplitudes from the EHT in 2009 \citep{Doeleman12} and 2012 \citep{Akiyama15}. Because the total compact flux density of M87 was similar for these two epochs, we include visibilities from both in the same plot. The corresponding visibilities obtained from the fiducial 230~GHz snapshot images (Fig.~\ref{fig::im230}) for model \texttt{H10} and \texttt{R17} are displayed by red and blue circles, respectively. At the chosen values of inclination $\theta=17^\circ$ and distance to the black hole $D=16.7$~Mpc, the images from the simulations are somewhat too large and, therefore, underpredict the measured visibility amplitudes on the Hawaii-California and Hawaii-Arizona baselines. At a larger inclination angle $\theta=30^\circ$, sampled visibility amplitudes from the simulated images (denoted by red and blue \texttt{x}s) better match the observations. 
}
\label{fig::vis}
\end{figure}

\begin{figure*}
\centering
\includegraphics*[width=0.99\textwidth]{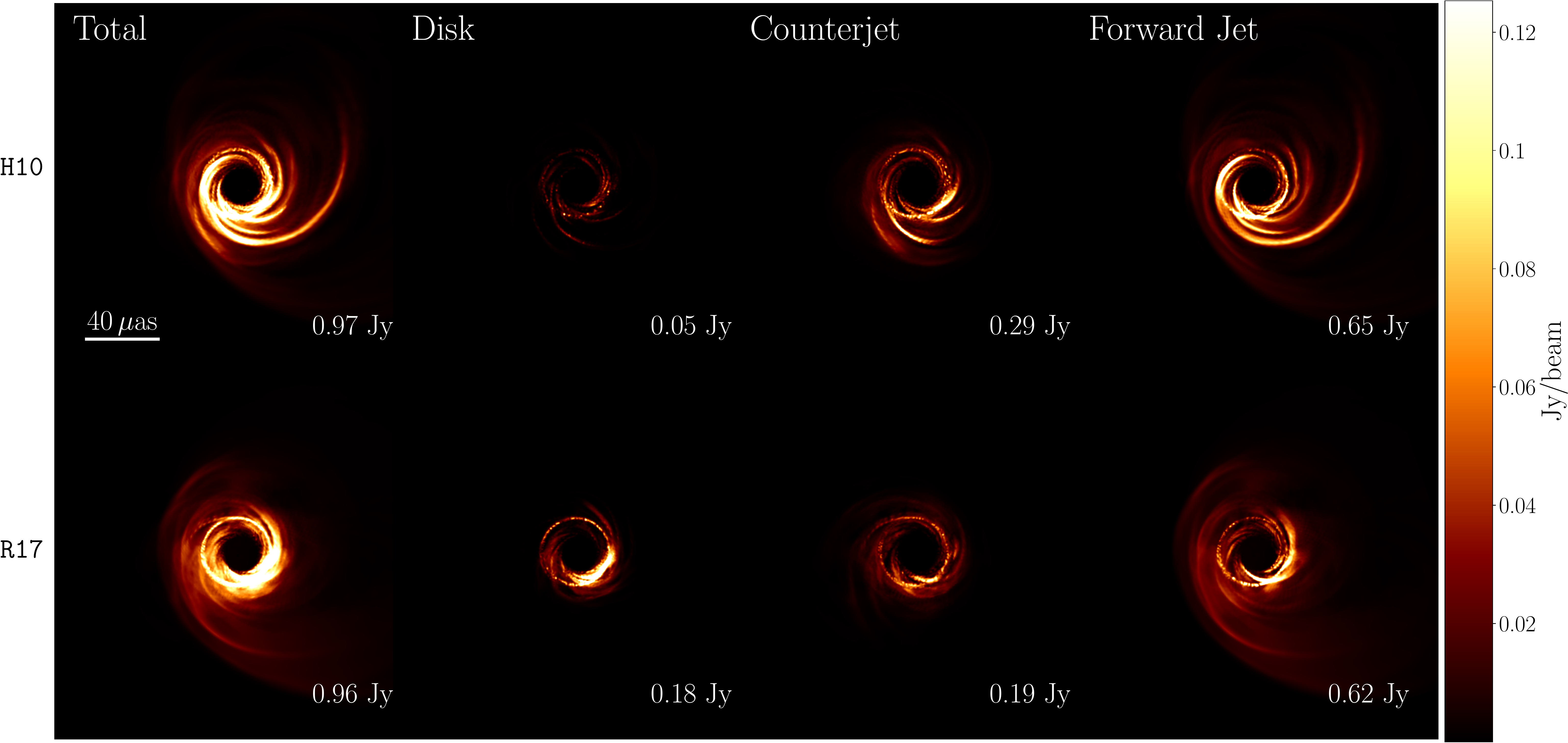}
\caption{
Snapshot images of the two simulations generated using \texttt{grtrans} to zero out emissivities from selected regions and highlight the emission from different components of the accretion flow. The leftmost column shows the image produced including all regions in the radiative transfer with $\sigma_{\rm i}<\sigma_{\rm cut}$; these are the same images as in the middle row of Fig.~\ref{fig::im230}. The second column from the left shows the image generated from the disc, setting emissivities in the jet regions (defined as $\text{Be}>0.05$) to zero. The third column shows emission only from the counterjet ($\text{Be}>0.05$ and polar angle $\theta<\pi/2$). The fourth column shows emission from only the forward jet ($\text{Be}>0.05$ and polar angle $\theta>\pi/2$). Since the accretion flow is optically thin at 230 GHz, the total flux densities of the component images nearly add up to the total flux density in the image generated from the entire emissivity distribution. Both simulations have images that are dominated by emission originating in the forward jet. \texttt{H10} has more counterjet emission, and \texttt{R17} has more disc emission. 
}
\label{fig::components}
\end{figure*}

To compare our models against existing images and data from VLBI observations, we computed high resolution images at 15, 22, 43, 86, 230, and 345 GHz using \texttt{grtrans}. We only include emission from the jet out to $3000 \, r_{\rm g}$. With the small inclination angle of $17^\circ$ \citep{Mertens2016,Walker2018}, this corresponds to a maximum projected jet length in our images of $850 \, r_{\rm g}$, or ${\approx} 3$ mas. In contrast, jet emission at 43~GHz extends out to at least 20~mas \citep{Walker2018}. We chose a representative snapshot for each model where the core flux density at 230 GHz was close to the measured value of 0.98 Jy from the EHT \citep{Doeleman12, Akiyama15}. 

In Fig.~\ref{fig::logimages}, we show log-scale images of both models at 43, 86, and 230 GHz, each with a dynamic range of $10^4$. The jet structure is similar in both simulations, with a wide apparent opening angle that increases to $>90^\circ$ at the jet base in the 230 GHz image. The jets in both models show filamentary magnetic field structure close to the black hole that rotates clockwise as viewed from our selected orientation. The spiral filaments are more prominent in the \texttt{H10} snapshot. In general, emission in the \texttt{H10} model comes from the high-temperature, high-magnetic field inner jet, and magnetic filaments in the jet dominate over disc emission at 230 GHz (Fig.~\ref{fig::components}). At longer wavelengths, the jet images produced from the two models are qualitatively similar, but the higher jet power in the \texttt{R17} model leads to brighter emission at larger distances from the black hole at all frequencies, while the core is consistently brighter in the \texttt{H10} images.

\subsubsection{Core-Shift}
\label{sec:core_shift}
VLBI observations with absolute phase referencing allow estimates of the relative image location at different frequencies. Because the M87 jet is optically thick at wavelengths longer than a few millimeters, the image centroid of the bright, compact core emission moves with frequency, giving rise to the so-called ``core-shift'' effect \citep{BK79}. 

\citet{Hada2011} conducted measurements of the core-shift of M87 at frequencies from 2.3 to 43~GHz, finding that the millimeter core is coincident with the SMBH and disc that launch the jet.  They estimated that the radio core has a right ascension displacement (relative to the 43~GHz core) given by $\Delta {\rm RA} \approx A \lambda_{\rm GHz}^{-\alpha} + B$, where $A=(1.40 \pm 0.16)\,{\rm mas}$, $B=(-0.041\pm 0.012)\,{\rm mas}$, and $\alpha=0.94\pm 0.09$. We computed the analogous core-shifts for our simulated images by first convolving the images with the wavelength-dependent observing beam of \citet{Hada2011} and then measuring the location of the peak in the resulting image. 

Fig.~\ref{fig::core_shift} shows the results of this analysis for our simulations. Both \texttt{H10} and \texttt{R17} produce images with core-shifts that are compatible with the results of \citet{Hada2011} at frequencies as low as 15~GHz. Even though VLBI constrains the core-shift at yet lower frequencies, we did not attempt to estimate core-shifts at these frequencies from our simulations because the image sizes and observing beams are comparable to the raytracing domain. 

Radio-jet core-shifts can be used to measure the jet magnetic field. Recently, \citet{Zaman2014} and \citet{Zdziarski2015} used core-shifts to measure the jet magnetic fields of several LLAGN sources (including M87) on $\sim$pc scales; they found their values of magnetic field and jet powers were consistent with jets launched by MADs. This is consistent with the findings of this work for M87. 

The magnetic field strength can also be estimated using the measured core sizes from VLBI. \citet{Kino2014} used this method with a model of the optically thick 43 GHz synchrotron emission to estimate a field strength 1 G $\lsim |B| \lsim $ 15 G at an angular scale of $100\,\mu$as. Assuming an inclination angle of $17^\circ$, an angular scale of $100\mu$as corresponds to $r\approx 100 r_{\rm g}$. From the time- and azimuth-averaged simulation data, we find for both models 1 G $\lsim |B|_{100\,r_{\rm g}} \lsim $ 2 G, within the measured range. Closer to the black hole, \citet{Kino2015} used EHT data to estimate a field strength 58 G $\lsim |B| \lsim $ 127 G on scales $\sim10\mu$as, corresponding to a de-projected distance of ${\approx}10\,r_{\rm g}$.  From our averaged simulation data, we estimate  $|B|_{10r_{\rm g}} \approx $ 20 G at this radius in the jet. However, \citet{Kino2015} obtain their estimate by assuming a spherical 230 GHz emission region that is optically thick to synchrotron self-absorption, which we do not observe at 230 GHz in our simulations. 

\subsubsection{43 and 86 GHz Images}
\label{sec::43}
VLBI images show that the jet of M87 is wide, with an apparent opening angle that decreases with radius. The apparent opening angle is ${\approx} 55^\circ$ at 43 GHz \citep{Walker2018}, increasing up to ${\approx} 127^\circ$ in the innermost regions $({\sim}50 r_{\rm g})$ of 86 GHz images \citep{KimM87}.  Fig.~\ref{fig::im43} compares the simulation images of the inner $2$ mas at 43 GHz with an image of 2007 VLBA data  \citet{Walker2018} reconstructed with a new closure phase and amplitude imaging method implemented in the \texttt{eht-imaging} software library (Fig. 9 of \citealt{Chael18_Closure}). The top row shows the snapshot images for both simulations, and the bottom row shows the VLBI image and the snapshot images convolved with a beam that has a size half of the nominal value reported in \citet{Walker2018}.  Lines indicating the $55^\circ$ apparent opening angle measured by ridge line analysis in \citet{Walker2018} are overlaid on all images. Previous simulations of M87 using weakly magnetized discs produced narrow jets \citep{Moscibrodzka_16M87, RyanM87}, with opening angles ${\lsim} 30^{\circ}$. In contrast, the observed wide apparent opening angle is naturally produced in these MAD simulations. The simulation images blurred to the same resolution as the VLBI image also show limb brightening in the jet, though the contrast between intensity on the jet edges and along the axis is in general less prominent than in the VLBI image. The counterjet is faint but visible at the edge of our dynamic range in both model images, but it is more prominent in the VLBI image at this epoch.

Fig.~\ref{fig::im86} compares the simulation images at 86 GHz with the 2014 image reconstructed from GMVA observations in \citet{KimM87} (their fig. 3, panel d). Again, both the \texttt{H10} and \texttt{R17} simulations produce wide opening angle jets consistent with the \citet{KimM87} image. At these frequencies, the opening angle of \texttt{R17} is slightly larger than that of \texttt{H10}, likely due to the fact that this simulation is even more MAD (Table~\ref{tab::summary}).  Like the GMVA image, the 86 GHz model images also show noticeable counterjet emission. However, the limb brighting in the blurred simulation images is somewhat less than in the VLBI image. 

\subsubsection{230 GHz Images}
\label{sec::230GHz}
At 1.3 mm, The EHT has already constrained the emission size at the core to be on the order of ${\sim}40\,\mu$as \citep{Doeleman12, Akiyama15}, approximately the size of the lensed black hole shadow for $M=6.2\times10^9 \, M_\odot$ \citep{Gebhardt11} and $D=16.7$ Mpc \citep{Mei2007}. More recent EHT observations with the full array are expected to directly image the jet launching region and potentially the black hole shadow itself \citep{Lu2014, akiyama_M87Imaging}.

Fig.~\ref{fig::im230} compares three linear scale images at 230 GHz showing time evolution of the source over roughly 1 week, the usual length of observation campaigns by the EHT. In each set, the top row shows high-resolution images from \texttt{grtrans}, and the bottom row shows the same images blurred  with a Gaussian beam with a $15\,\mu$as full width at half maximum (FWHM). This beam size is  approximately the imaging resolution of the EHT \citep{Chael18_Closure}.

The 230 GHz images for both models show a distinct black hole shadow that is large enough to be imaged by the EHT. The diameter of the shadow is approximately $9.7 \, r_{\rm g} \approx 36 \, \mu{\rm as}$ (slightly less than the diameter for a Schwarzschild black hole, $2\sqrt{27}r_{\rm g}\approx38 \, \mu$as), given the assumed mass and distance. Because we do not view the accretion flow directly face-on, the bottom half of the ring structure is brighter due to Doppler boosting of the jet and disc emission. Based on the differential limb brightening and velocities in the large-scale jet \citep{Walker2018}, we chose the jet to rotate clockwise in the plane of the sky. Given the sense of rotation determined by \citet{Walker2018} and the direction of the projected jet on the sky, the 230 GHz ring is expected to be brighter on the bottom from Doppler boosting. 

Bright ridges tracing the rotating helical magnetic field are visible in both simulations. These extended structures are fainter than the bright ring, but they are visible even in the images blurred to the EHT's resolution. The brightest spot in the 230~GHz image moves with the rotation of the magnetic field lines, particularly in the \texttt{H10} model, which lacks bright disc emission. When blurred to EHT resolution, this evolution will generally follow the clockwise sense of rotation of the disc and jet. In the specific frames selected from the \texttt{H10} simulation, however, shifts in the relative brightness of two filaments as they rotate  produce an apparent evolution in the blurred frames that is slightly counterclockwise. 

The precise shadow diameter and shape is sensitive to the inclination and black hole spin \citep{Bardeen, Chandra83}. Even for these simulated images, which have a prominent shadow, any one EHT image reconstruction would leave substantial uncertainty in the shadow size due to the limited resolution and contributions to the source structure from the foreground jet. It may be possible, however, to make a more precise measurement of the shadow size with multi-epoch imaging. While the shadow is a persistent feature set only by the mass and spin of the black hole, the foreground  jet at 230 GHz rotates quickly, completing a full revolution on a time-scale of weeks to months. 

In Fig.~\ref{fig::vis}, we compare visibility amplitudes extracted from the fiducial 230 GHz images in Fig~\ref{fig::im230} with observations from the EHT with stations in Hawaii, California, and Arizona in 2009 \citep{Doeleman12} and 2012 \citep{Akiyama15}. The compact flux density measured in these two years was ${\approx} 0.98$ Jy. At the chosen values of inclination, $\theta=17^\circ$ up from the south pole, and distance to the black hole $D=16.7$ Mpc, the snapshot images from our simulations are somewhat too large and underpredict the measured visibility amplitudes on the Hawaii-California and Hawaii-Arizona baselines. Compared to the extremely compact images obtained from the less magnetized 2D simulations in \citet{RyanM87}, our wide-opening-angle jets produce extended emission that increases the overall image size, though the bright ${\sim}40\mu$as photon ring remains the most prominent feature. 

We note that the image size in our simulations is highly sensitive to the assumed viewing inclination. At larger values of inclination angle than the $\theta=17^\circ$ taken from \citet{Walker2018}, the image size decreases as we see less Doppler-boosted emission from the wide opening angle jet. While the jet inclination angle is constrained to ${\lsim}20^\circ$ at distances ${\sim}100$pc from the black hole from apparent superluminal velocities measured near the HST-1 knot \citep{Giroletti}, it is not as definitively constrained on scales closer to the black hole. In their conservative estimate, \citet{Mertens2016} give an upper limit $\theta \lsim 27^\circ$.  At $\theta=30^\circ$, the visibilities from \texttt{R17} nearly match the observations, and the simulated visibilities from \texttt{H10} are much less discrepant than at $17^\circ$ (Fig.~\ref{fig::vis}).  Furthermore, at this larger value of the inclination angle, the limb-brightening in the 43 and 86 GHz images from both simulations more closely matches the VLBI maps (left panels of Figs.~\ref{fig::im43} and~\ref{fig::im86}), and the counterjet is more prominent at both frequencies. 


The image size in our simulations at a fixed inclination angle is also sensitive to the $\sigma_{\rm i}$ cut that we impose in post-processing. Our choice of $\sigma_{\rm cut} = 25$ was determined to avoid including emission from the regions with densities set to the floor value. In nature, these regions will produce emission which is likely to be significant in the regions probed by the 230 GHz image. We find that including radiation from $\sigma_{\rm i}>25$ regions makes the image more compact (see Section~\ref{sec::sigcut}). Thus, it remains possible that different prescriptions for including radiation in post-processing from highly magnetized regions may produce images from MAD simulations that are consistent with the 2009 and 2012 EHT size measurements. 

In Fig.~\ref{fig::components}, we consider the same 230 GHz snapshots as in the central column of Fig.~\ref{fig::im230} and decompose the emission into three component parts: disc, counterjet, and forward jet. As above we define regions with $\rm{Be}>0.05$ to be in the jets, and $\rm{Be}\le0.05$ to be in the disc. We produce these component images by zeroing out the emissivities outside the selected regions when doing the radiative transfer in \texttt{grtrans}. Throughout, we maintain the global cut on emissivities from regions with $\sigma_{\rm i} > 25$. 

Because the entire M87 accretion flow is optically thin at 230 GHz, the total flux densities of the component images in Fig.~\ref{fig::components} nearly add up to the total flux density in the image generated from the entire emissivity distribution. In both simulations, the majority (${\sim}60$ per cent) of the emission comes from the forward jet. The forward jet emission in both cases consists of a spiral structure surrounding the black hole where emission traces magnetic field lines, as well as a persistent component from the photon ring. We note that a substantial fraction of the jet emission comes from between the $\rm{Be}=0.05$ and $\sigma_{\rm i}=1$ surfaces, so adopting $\sigma_{\rm i}>1$ as the definition of the jet region will result in assigning much more of what we classify as forward jet and counterjet emission to the disc.  

In the \texttt{H10} model, nearly all of the emission not from the forward jet is produced from the counterjet, which adds to the prominent photon ring. Because the electrons in this model are so much hotter at the base of the jet than in the disc, the emission from these regions dominates the total and the disc emission is negligible. In contrast, the reconnection heating model \texttt{R17} has approximately equal contributions from the counterjet and disc. The disc emission shows a persistent bright spot from the Doppler-boosted accretion flow. Unlike the bright spots produced from rotating magnetic field lines, this spot remains constant in position and does not rotate around the photon ring with time. This feature points to the possibility of using multi-epoch imaging with the EHT to disentangle the source structure and identify whether or not the accretion disc emission makes a substantial contribution to the image flux. 

\subsection{The effects of our choice of $\sigma_{\rm cut}$}
\label{sec::sigcut}

\begin{figure*}
\centering
\includegraphics*[width=0.9\textwidth]{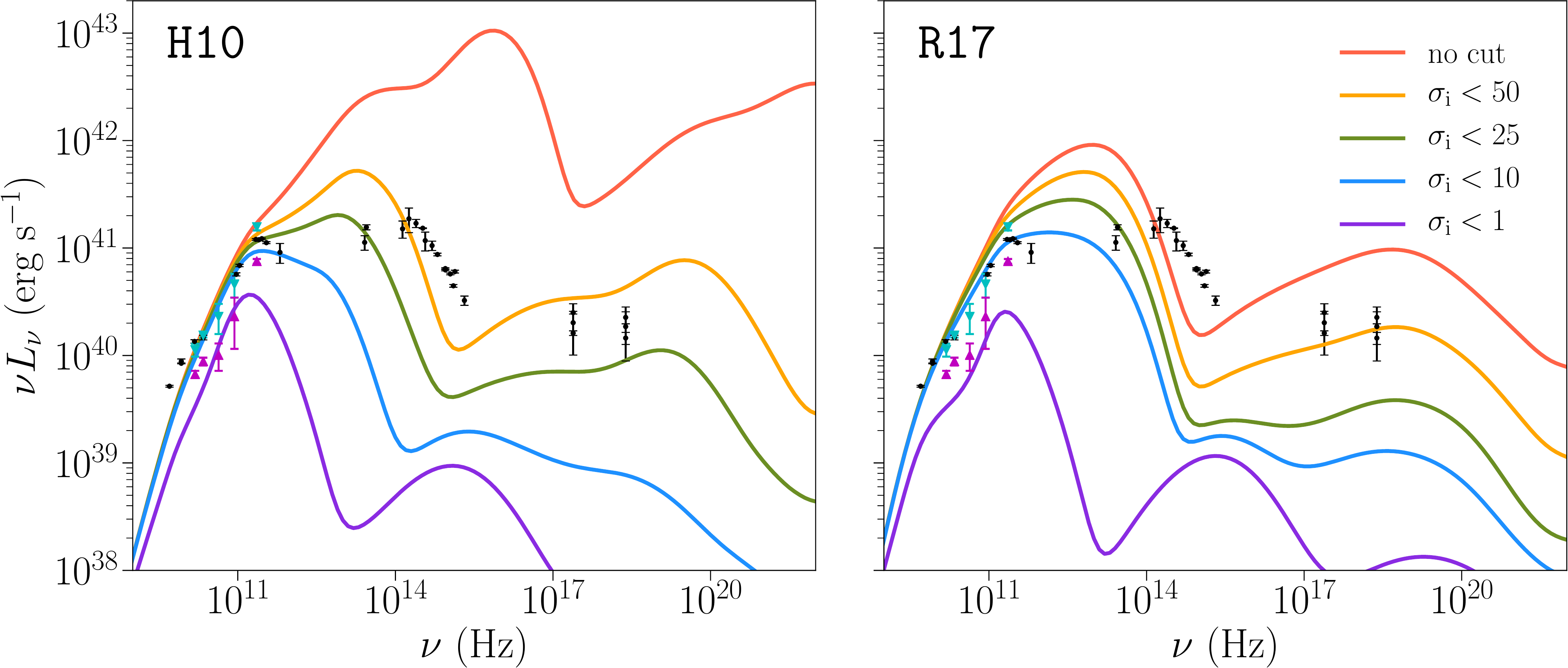}
\caption{
Snapshot spectra from the two simulations generated with different values of $\sigma_{\rm cut}$ in the radiative transfer. Spectra were generated with \texttt{HEROIC} zeroing out emissivities from all regions with fluid frame magnetization $\sigma_{\rm i}\ge\sigma_{\rm cut}$. We generate spectra with $\sigma_{\rm cut}=$1, 10, 25 (our fiducial value), 50, and with no ceiling. Because the images have different total flux densities, the intensity scale at each $\sigma_{\rm cut}$ is different.
In both simulations, any choice of $\sigma_{\rm cut}$ above unity has little effect on the radio spectrum up to 230 GHz. Most emission in this part of the spectrum comes from less-magnetized regions farther from the black hole. The choice of $\sigma_{\rm cut}$ has a drastic effect on the spectrum at higher frequencies as direct synchrotron emission and Compton scattering in the most magnetized, high-temperature regions close to the black hole is added, increasing the radiative power. When no $\sigma_{\rm i}$ ceiling is imposed, model \texttt{H17} has an extreme total luminosity $>10^{43}$ erg s$^{-1}$.
}
\label{fig::sigcut_spec}

\centering
\includegraphics*[width=\textwidth]{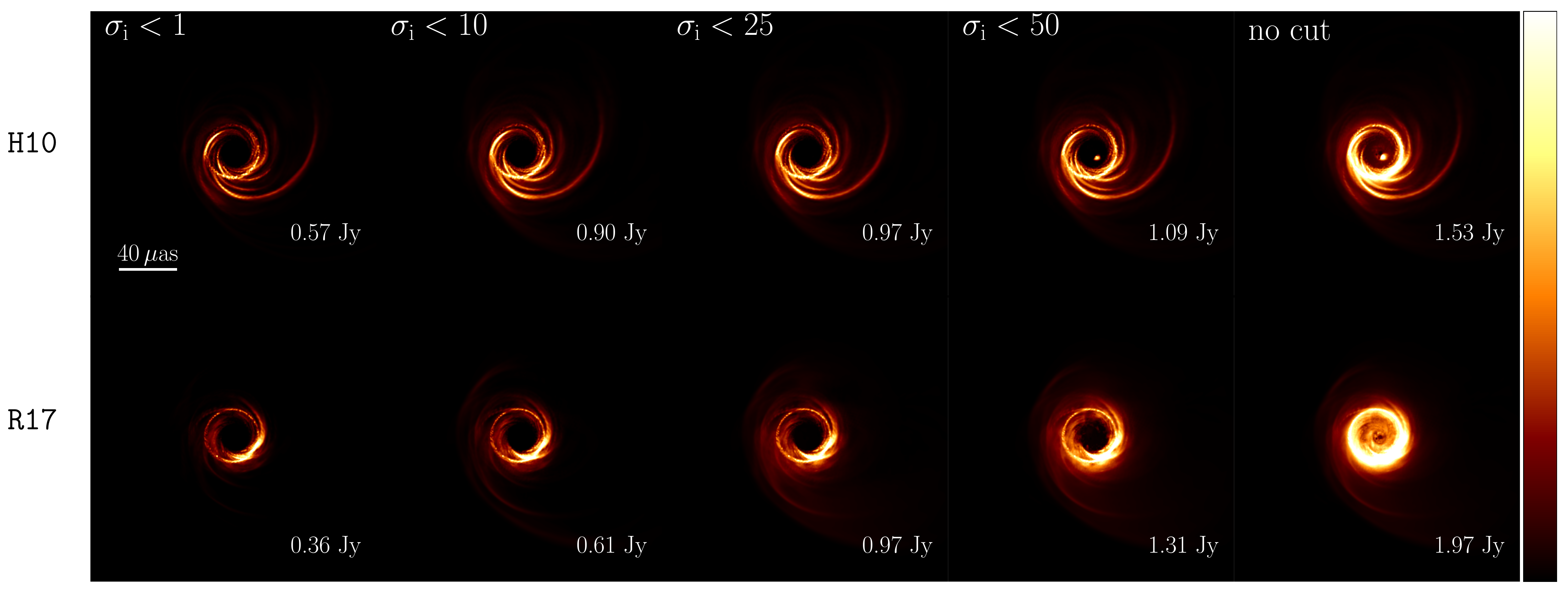}
\caption{
Snapshot images from the two simulations generated with different values of $\sigma_{\rm cut}$ in the radiative transfer. From left to right, we present images generated using $\sigma_{\rm cut}=$1, 10, 25 (our fiducial value), 50, and with no ceiling.  In both simulations, the overall image structure is similar at all cuts up to $\sigma_{\rm cut}=50$. Because $\sigma_{\rm i}$ increases rapidly with decreasing polar angle in the jet region (Fig.~\ref{fig::symc2}), including regions of higher and higher magnetization does not open up very different regions of the accretion flow to the radiative transfer. In contrast, including the entire interior of the jet (the rightmost images) produces substantial new emission in front of the photon ring.
}
\label{fig::sigcut}
\end{figure*}

In this section, we explore the effects of our choice of choosing $\sigma_{\rm cut} = 25$ in the results presented in the previous sections. As discussed in Section~\ref{sec::reliability}, a choice of $\sigma_{\rm cut}$ is necessary in radiative transfer in order to exclude emission from the regions that most evidently suffer from errors in the gas-dynamical evolution. Although the ratio of electron-to-ion temperatures behaves as we expect from our two heating prescriptions in this region (Fig.~\ref{fig::sym1}), the overall temperature scale from the gas evolution may suffer from errors in any region with $\sigma_{\rm i}>1$. On the other hand, a value of $\sigma_{\rm i} > 1$ may be necessary to investigate jet emission from these simulations and compare to data. 

For our chosen simulation snapshots, we test the effects $\sigma_{\rm cut}$ by  re-generating spectra and 230 GHz images testing five different values of $\sigma_{\rm cut}$: $\sigma_{\rm cut}=$1, 10, 25 (our fiducial value), 50, and no ceiling. In Fig.~\ref{fig::sigcut_spec}, we see that in both simulations, any choice of $\sigma_{\rm cut}>1$ has relatively little effect on the radio spectrum up to 230 GHz. Most emission in this part of the spectrum comes from less-magnetized material with $\sigma_{\rm i} < 25$ farther up from the black hole. As a result, the predictions of both models at frequencies $<$ 230 GHz should be relatively insensitive to the choice of $\sigma_{\rm cut}$. 

For $\sigma_{\rm cut} = 1$, our simulations are under-luminous across the radio spectrum. This change in the luminosity with decreasing $\sigma_{\rm cut}$ from 10 to 1 indicates that the majority of the radio flux originates in regions in the jet with $1<\sigma_{\rm i}<10$. However, this inference is dependent on our overall normalization of our simulations chosen to match the observed 230 GHz flux density with $\sigma_{\rm cut}=25$. 

For $\nu\gsim$ 230 GHz, direct synchrotron emission and Compton scattering in the most magnetized, high-temperature regions close to the black hole makes a substantial contribution to the radiative power. The predictions of our models at these frequencies are thus strongly dependent on the choice of the $\sigma_{\rm cut}$, and should be viewed with caution. 
Although the imposition of a global ceiling on $\sigma_{\rm i}<\sigma_{\rm i,max}$ for stability when running the simulation makes it impossible to conclusively predict the emission from our models at these higher frequencies, we observe some trends in the spectra with increasing $\sigma_{\rm cut}$ in Fig.~\ref{fig::sigcut_spec} that should be explored in future work. In model \texttt{R17}, the overall luminosity, peak synchrotron frequency, and Compton power all increase when we add increasingly magnetized regions of the simulation, but the overall spectral shape does not drastically change. It seems possible that in future simulations with a higher absolute bound on $\sigma_{\rm i,max}$, we could extend the ceiling on $\sigma_{\rm cut}$ imposed in the radiative transfer and produce a model that better matches the higher frequency measurements. 

In contrast, when no $\sigma_{\rm cut}$ is imposed, model \texttt{H17} has an extremely large total luminosity $>10^{43}$ erg s$^{-1}$, dramatically exceeding measurements at all frequencies above 230 GHz. As we saw in Section~\ref{sec::properties}, this extreme luminosity affects the dynamics of the fluid, most notably by reducing the mechanical jet power in this model relative to \texttt{R17}. This extreme luminosity ultimately results from the high electron temperatures produced by $\delta_{\rm e}\rightarrow 1$ in the most highly magnetized regions near the black hole. However, because the density in these regions is set by the $\sigma_{\rm i,max}=100$ ceiling, it remains possible that a simulation using the \citet{Howes10} turbulent cascade heating prescription with a higher value of $\sigma_{\rm i,max}$ and a correspondingly more evacuated jet would produce more reasonable total radiative power.

In Fig.~\ref{fig::sigcut}, we present 230 GHz images from the two simulations using different choices of $\sigma_{\rm cut}$. In both simulations, the overall image structure is similar at all cuts up to $\sigma_{\rm cut}=50$. Since $\sigma_{\rm i}$ increases rapidly with polar angle in the jet region (Fig.~\ref{fig::symc2}), including regions of higher and higher magnetization does not open up very different regions of the accretion flow to our radiative transfer as long as we remain below the overall simulation ceiling. 

In the rightmost images in Fig.~\ref{fig::sigcut}, we do not impose a ceiling on $\sigma_{\rm i}$ in the radiative transfer, opening up the entire interior of the jet. This produces substantial new emission from both simulations that originates from close to the black hole, dramatically increasing the brightness and compactness of the images. In the \texttt{H10} model, the forward jet compact emission is concentrated in a bright spot in the middle of the ring from the very-high-temperature electrons at the jet base, while in the \texttt{R17} model, the highest-$\sigma_{\rm i}$ emission forms a more diffuse haze in the middle of the jet in front of the photon ring. Furthermore, in both models, the addition of strong emission close to the black hole dramatically increases the prominence of the lensed counterjet emission in the photon ring, to the point where the forward jet and counterjet contributions to the 230 GHz image become approximately equal. The strength of the emission at the base of the jet/counterjet sets the relative brightness of the photon ring to the surrounding disc and jet \citep{Dexter12}, but this emission is unfortunately unconstrained in our models because of the uncertainties associated with the choice of $\sigma_{\rm cut}$. 

\begin{figure}
\centering
\includegraphics*[width=0.45\textwidth]{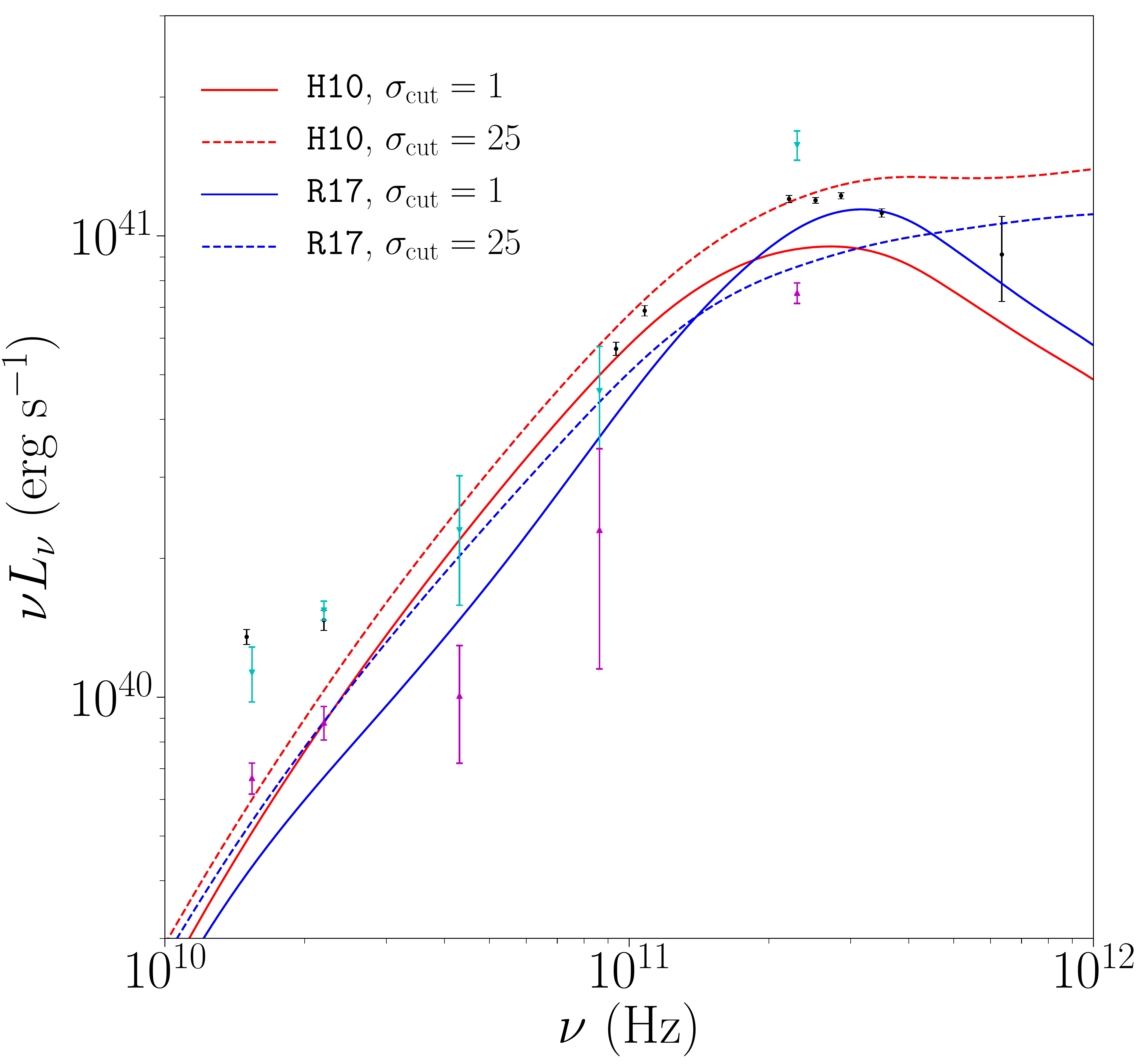}
\caption{
Median synchrotron radio spectra from the two simulations generated with \texttt{grtrans} testing two different combinations of $\sigma_{\rm cut}$ and density rescaling in post-processing. The dashed red and blue lines show the \texttt{H10} and \texttt{R17} synchrotron spectra computed with $\sigma_{\rm cut}=25$ and no density rescaling, as in Fig.~\ref{fig::spectra}. The solid red and blue lines show the corresponding spectra computed with $\sigma_{\rm cut}=1$, after rescaling the simulation density (and energy density/magnetic field) by a factor of 1.2 for (\texttt{H10}) and 2.2 for (\texttt{R17}). The radio spectra from the rescaled simulations limited to $\sigma_{\rm cut}=1$ match the slope of our $\sigma_{\rm cut}=25$ fiducial spectra, but turn over much more rapidly at the peak around 230 GHz. 
}
\label{fig::spec_scale}
\end{figure}

Finally, in Fig.~\ref{fig::spec_scale}, we test the necessity of including the $25>\sigma_{\rm i}>1$ jet emission by re-computing spectra using $\sigma_{\rm cut}=1$, re-scaling the density and magnetic field up in postprocessing to match ${\approx}0.98$Jy at 230 GHz. Model \texttt{R17} was scaled by a factor of 2.2, and \texttt{H10} was scaled by 1.2. The radio spectra from the rescaled simulations limited to $\sigma_{\rm cut}=1$ match the slope of our $\sigma_{\rm cut}=25$ fiducial spectra, but turn over much more rapidly at the peak around 230 GHz. Fig~\ref{fig::spec_scale} indicates the choice of $\sigma_{\rm cut}$ (combined with an appropriate density normalization) is an important free parameter in the postprocessing of spectra from MAD simulations. Because rescaling the simulations in postprocessing can affect the radiative efficiency \citep[which is non-negligible for M87][]{RyanM87}, we do not show further results from these rescaled spectra. 

\section{Discussion: Comparison to Previous models}
\label{sec::discussion}
The two MAD simulations presented in this paper produce spectra and images that are broadly consistent with observations of the M87 jet at centimetre and millimetre wavelengths. Model \texttt{R17} produces a jet power that is in line with observations, while model \texttt{H10} produces a jet power lower by a factor of two, unless radiation (made suspect by its origin in regions with active density and energy floors) is considered. Both models reproduce the wide jet opening angle observed at 43~GHz, and they are both consistent with core-shift observations. However, at a viewing angle of $17^\circ$, the 230 GHz images from both models are too large to match EHT observations from 2009 and 2012.

M87 has been simulated before using GRMHD and GRRMHD codes, though not as frequently as \sgraa. The model of \citet{Moscibrodzka_16M87} is a representative state-of-the-art combination of single-fluid GRMHD and radiative transfer. The authors used discs with relatively weak magnetization from \citet{Shiokawa} and added thermal electrons in post-processing that were assumed to be hot in the jet and cool in the disc \citep{Moscibrodzka_14}. They produce a model with a jet power in the correct range, a flat radio spectrum, and an limb-brightened jet image at 43 GHz. The jets in their simulations show substantial variability and apparent superluminal motion from field lines along the funnel wall. However, the jets in their simulation have apparent opening angles at 43 GHz smaller than the observed ${\sim}55^\circ$. 

As in the MAD models in this work, 230 GHz images from \citet{Moscibrodzka_16M87} show spiral structures from helical field lines in the jet; this is a common prediction of both weakly magnetized and MAD models. At 230 GHz, their images are dominated by the counterjet \citep[see also][]{Dexter12}. Unlike in our MAD models, their 230 GHz images satisfy the constraints on the image size from the EHT at $20^\circ$ inclination.

The authors have also investigated the polarized emission from their models \citep{Moscibrodzka_18}. They have shown that in their counterjet-dominated models it is possible to produce images with rotation measure and polarization fraction in line with observations through the depolarization of the counterjet emission as it passes through the cooler disc. The emission in both of our MAD models at 230 GHz is dominated by the forward jet, although counterjet emission is still significant. It is possible that, with less opportunity for depolarization through the disc, our forward-jet-dominated images might produce a net polarized flux that exceeds the observed value (${\sim}1$ per cent; \citealt{Kuo2014}). This is an important direction for future work.

Recently, \citet{RyanM87} preformed the first two-temperature simulation of M87 using the code \texttt{ebhlight} in axisymmetry. Unlike \texttt{KORAL}, which considers only the frequency-integrated radiation field, \texttt{ebhlight} uses a Monte Carlo method where photons with distinct frequencies are emitted and absorbed on the simulation grid. Consequently, they obtain spectra as a natural product of their simulations without having to perform radiative transfer in post-processing. 

\citet{RyanM87} considered discs that were far less magnetized than those explored here. Consequently, to match the observed 230 GHz flux density they required higher accretion rates than in our simulations. In their best-fitting model, the accretion rate is $\dot{M}/\dot{M}_{\rm Edd}\sim10^{-6}$, about $3\times$ what we find in our lower-accretion-rate model \texttt{H10}. In all their models, they found that Coulomb heating of electrons becomes important in the outer disc. As in our simulations, they see that radiation plays a significant role in the inner disc.  Notably, they explore simulations with both high \citep{Gebhardt11} and low \citep{Walsh13} black hole mass and consider two values of the black hole spin ($a=0.5$ and $a=0.9375$). They find their high-spin, high mass model produces both a spectrum and 230 GHz image consistent with the available data -- we have adopted these preferred parameter values in this study. Unlike in the present work, at $M=6.2\times10^9 M_\odot$ and $a=0.9375$, they obtain a compact, counterjet-dominated 230 GHz image that is consistent with past EHT measurements of the overall image size. However, the jet powers produced in their simulations are several orders of magnitude too low, and their weakly magnetized discs also produce jet opening angles that are too small when compared against VLBI observations.  These problems are not present in our MAD models, which match the observed spectral and image characteristics of the M87 jet well at all frequencies between 15 and 86~GHz.

Simulating the jet interior remains a problem in all simulations, not just in the MAD regime, and all simulations must impose some sort of density floor in the magnetized, evacuated jet to ensure numerical stability. As discussed in Section~\ref{sec::sigcut}, this problem is particularly important in MAD models where much of the emission may come from these highly magnetized regions. The matter content of the jet is still unknown; it may be populated by a pair plasma of electrons and positrons \citep{Moscibrodzka_11, Broderick15}. Further work with our simulations using additional postprocessing prescriptions for the jet matter content and electron distribution may be able to provide constraints on models for the jet interior, while still relying on predictions from self-consistent temperature evolution in the jet wall and disc regions of the simulation.

\section{Summary and Conclusion}
\label{sec::summary}
We have presented in this paper the first Magnetically Arrested Disc simulations performed with two-temperature, radiative, general relativistic magnetohydrodynamics. After normalizing to the 230 GHz flux density, the two simulations \texttt{H10} and \texttt{R17} produce spectra consistent with that of M87 in the radio, millimetre, and submillimetre. They both produce powerful jets with jet power consistent with or close to the measured range. The jets both have a wide opening angle consistent with the wide opening angle of the M87 jet observed with VLBI at 43 and 86 GHz. The simulated images at centimetre and millimetre wavelengths exhibit a core-shift which reproduces that reported in \citet{Hada2011}. The 230~GHz images from both simulations clearly show the lensed photon ring, i.e. the black hole shadow, indicating that even in forward-jet-dominated MAD images, the full EHT should be able to image this feature. In addition, the images are dynamic on time-scales of months to years. If these images are reflective of M87, then repeated EHT observations should be able to detect the motion of rotating magnetic fields that are driven by the spin of the black hole. 

Two sub-grid electron heating mechanisms were  considered in this work, and they produce jet outflows with somewhat distinct properties. The magnetic reconnection heating model of \citet{Rowan17,Chael18} used in simulation \texttt{R17} launches a jet powered by the black hole spin with a mechanical jet power ${\sim}10^{43}$ erg s$^{-1}$, in the correct range for M87 \citep{Reynolds96_2,Stawarz06}. In contrast, the jet heated by the turbulent cascade model of \citet{Howes10} in simulation \texttt{R17} produces intense radiation at the jet base; this radiation saps the jet of mechanical energy, resulting in a mechanical jet power a factor of two less than in \texttt{R17}. 

Despite the differences in their kinematics, the spectra and images of the two models are quite similar at centimetre, millimetre, and submillimetre wavelengths. The simulations did not run long enough to produce the full extent of the jet observed with VLBI, but the images from  the inner milliarcsecond show similar structure to VLBA and GMVA images at the corresponding wavelengths. Notably, both models reproduce the measured ${\approx}55^\circ$ opening angle \citep{Walker2018}, and both show emission from the counterjet, though our models produce less counterjet emission than observed in some 43~GHz VLBI images.

At 230~GHz, our simulations produce images that are larger than the size measured by the EHT in 2009 and 2012 \citep{Doeleman12, Akiyama15}. However, the image size is strongly dependent on the viewing inclination, and it becomes consistent with EHT observations around $\theta=30^\circ$, the upper end of the most conservative range established by VLBI observations \citep{Mertens2016}. The image size also shrinks when the extremely magnetized on-axis regions we exclude in our radiative transfer are included. In this work, we do not trust the emission from these most magnetized regions of our models due to the imposition of density floors; we exclude emission from regions with a magnetization greater than $\sigma_{\rm cut}=25$. A different treatment of these regions from a simulation with a higher overall ceiling on the magnetization could potentially bring the size of the 230 GHz images in line with observations.

Our 230~GHz images all show distinct black hole shadows, which are primarily illuminated by emission originating in the forward jet. Future investigations of polarized images from our simulations will investigate whether our forward-jet-dominated models satisfy constraints on the 230 GHz polarization fraction; these constraints are naturally satisfied by Faraday depolarization of counterjet emission \citep{Moscibrodzka_18}. Furthermore, the rotation of the accretion disc and jet makes our 230 GHz images dynamic on time-scales of weeks to months. It may be possible to distinguish between models by identifying jet, counterjet, and disc contributions to the 230~GHz image by tracking stationary and moving features through repeated observations with the EHT. 

While our models under-produce the measured flux in the optical, ultraviolet, and X-ray, it seems clear that the spectrum at these high frequencies is dominated by the hottest, most magnetized regions closest to the black hole. A different treatment of the magnetization ceilings imposed in the simulation and radiative transfer may bring the simulation spectra from these regions more in line with observations. At the other end of the spectrum, we are unable to investigate the jet on large scales and at frequencies $<15$ GHz given the relatively short simulation time considered in this work.

Finally, we note that it is likely that nonthermal electrons in the disc \citep{Broderick09} or jet \citep{Dexter12} contribute to the emission at 230 GHz and into the infrared, optical, and ultraviolet. It is relatively straightforward to add non-thermal electron distributions to GRMHD simulations in postprocessing \citep{Dexter12, Ball_16, Davelaar_17}. However, the present work, as well as previous work by \citealt{Ressler17},\citet{RyanM87}, and \citet{Chael18}, has shown that including self-consistent evolution of electron populations in simulations substantially modifies the resulting images and spectra in ways that cannot be captured by simply assigning electron temperatures and distribution functions in post-processing. In this spirit, the logical next step is to self-consistently evolve non-thermal electron distributions in simulations of M87. In future work, we will use the method developed in \citet{Chael17} to this end.   

\renewcommand\thefigure{A.\arabic{figure}}
\setcounter{figure}{0}    
\begin{figure*}[t]
 \centering
 \includegraphics*[width=0.99\textwidth]{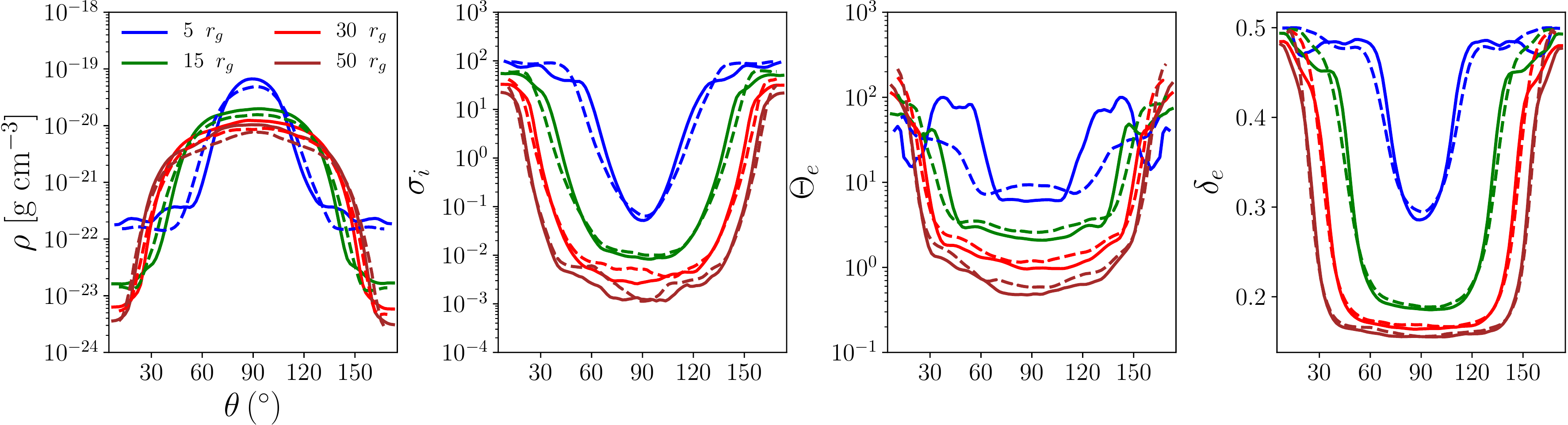}
 \caption{Quantities averaged over azimuth and time (from $t=10,00\,t_{\rm g}$ to $t=15,000\,t_{\rm g}$) from the original \texttt{R17} simulation (solid curves) and the lower resolution test run (dashed curves). The quantities are plotted at four radii: $r=5 \, r_{\rm g}$ (blue), $r=15 \, r_{\rm g}$ (green),  $r=30 \, r_{\rm g}$ (red), and $r=50 \, r_{\rm g}$ (brown). From left to right the plotted quantities are the gas density $\rho$, the magnetization $\sigma_{\rm i}$, the dimensionless electron temperature $\Theta_{\rm e} = k_{\rm B}T_{\rm e}/m_{\rm e}c^2$, and the electron heating fraction $\delta_{\rm e}$. 
 }
 \label{fig::rescompare_dists}
\end{figure*}

\begin{figure}
 \label{fig::rescompare_spec}
 \centering
 \includegraphics*[width=0.45\textwidth]{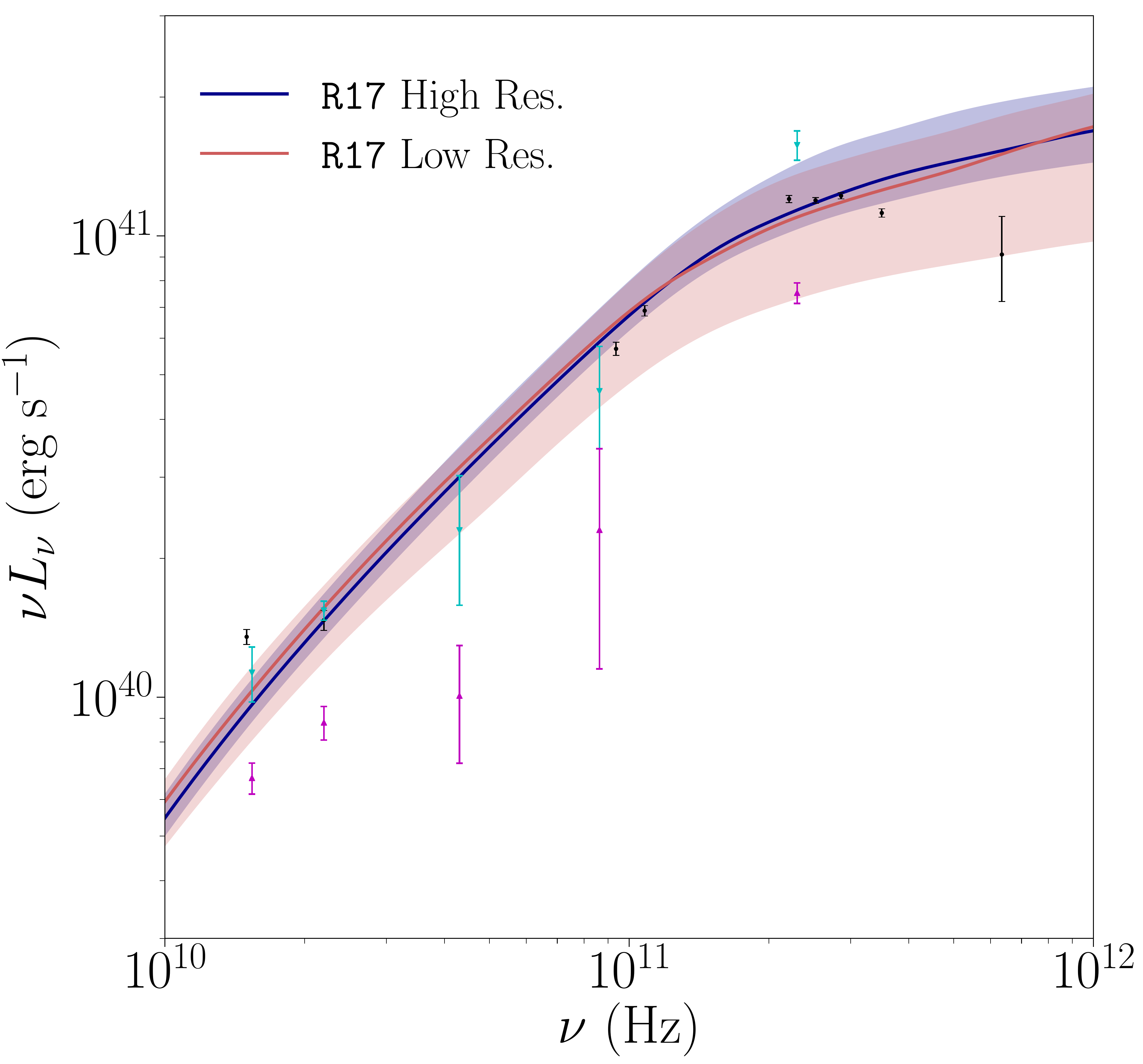}
 \caption{Synchrotron spectra from \texttt{R17}  for both the original high resolution (dark blue) and the low resolution test (light red) runs. Synchrotron spectra were computed over the range $10,000-15,000\,t_{\rm g}$ using \texttt{grtrans}; the solid lines indicate the median values, and the bands represent nominal $1\sigma$ variability ranges. The synchrotron spectra in the high and low resolution runs are converged to the same shape and level, but with more variability on the low resolution spectrum.  
 }
\end{figure}

\appendix 
\section*{Acknowledgements}
We thank the anonymous referee for their thorough and helpful comments, which have significantly strengthened this work. We thank Jason Dexter for insightful discussion of radiative transfer postprocessing choices and for his help in using and modifying \texttt{grtrans}. We thank Craig Walker for providing the 43 GHz VLBA data used in Fig.~\ref{fig::im43} and Jae-Young Kim for providing the 86 GHz GMVA image in Fig.~\ref{fig::im86}. We also particularly thank Monika Mo{\'s}cibrodzka and Ben Ryan for several productive conversations about previous simulation work on M87. 

This work was supported in part by NSF grant OISE-1743747. 
AC was supported in part by NSF grants AST-1440254 and AST-1312651. 
MJ was supported by NSF grant AST-1716536.
The authors acknowledge computational support from NSF via XSEDE resources (grant TG-AST080026N). 
This work was conducted at the Black Hole Initiative at Harvard University, 
supported by a grant from the John Templeton Foundation.

\setcounter{secnumdepth}{0}
\section{Low Resolution Test}

To test whether quantities of interest and the raytraced spectra from our simulations are converged, we repeated simulation \texttt{R17} using a lower resolution grid of $192\times128\times64$ cells in radius, polar angle, and azimuthal angle, respectively. We found this resolution to be near the minimum for which stable evolution of the rapidly changing magnetic field near the horizon was possible, using our original choices for the numerical floors. 
We used the same initial conditions, boundary conditions, $\sigma_{\rm i}$ ceiling, and rescaling strategy at $10,000\,t_{\rm g}$ as in the original, high-resolution run. 

Fig.~\ref{fig::rescompare_dists} shows comparisons of the average distributions of the density $\rho$, magnetization $\sigma_{\rm i}$, dimensionless electron temperature $\Theta_{\rm e} = k_{\rm B}T_{\rm e}/m_{\rm e}c^2$, and electron heating function $\delta_{\rm e}$ as a function of polar angle at four radii. The data were averaged in azimuth and over the time period between $10,000\,t_{\rm g}$ and $15,000\,t_{\rm g}$. The angular curves of average density, magnetization, and electron heating function show excellent agreement at all radii between the high and low resolution simulations. The electron temperature also shows excellent agreement at almost all radii and angles, but close to the black hole ($\lesssim5 \, r_{\rm g}$), the average electron temperature differs by up to a factor of four  between the high and low resolution runs at certain polar angles in the corona. The magnetization in this region is increasing rapidly with angle up to the maximum allowed value of $\sigma_{\rm i}=100$.

While the qualitative picture of the electron-to-ion temperature ratio induced by the heating prescription $\delta_{\rm e}$ is consistent between the two resolutions, this test indicates the precise values of the electron temperature close to the black hole may be resolution-dependent, as the effects of grid size on the computed dissipation in the highest $\sigma_{\rm i}$ regions becomes significant. Care should be taken to assess convergence before fitting exact values from simulation spectra or images to data.

Fig.~\ref{fig::rescompare_spec} shows synchrotron spectra of the high and low resolution \texttt{R17} models. In both cases, the spectra were computed with \texttt{grtrans} at 163$^{\circ}$ inclination using $\sigma_{\rm cut}=25$, with no additional density rescaling. The variability amplitude of the models changes with resolution, with the lower resolution simulation exhibiting stronger variations. However, the median spectrum in the high and low resolution runs is effectively the same, with excellent agreement between $10^{10}$ and $10^{12}$ Hz. This test indicates that the average simulation properties and the synchrotron spectrum at the frequencies of interest to this paper ($\nu < 10^{12}$ Hz) are converged in resolution.
The characteristics of the light-curve variability, on the other hand, are not converged with resolution, limiting our interpretation of the  results of Section~\ref{sec::variability}.   

\bibliography{ElecEv}

\end{document}